\documentclass[fleqn,usenatbib]{mnras}
\usepackage{fix-cm}
\usepackage{newtxtext,newtxmath}

\usepackage[T1]{fontenc}

\DeclareRobustCommand{\VAN}[3]{#2}
\let\VANthebibliography\thebibliography
\def\thebibliography{\DeclareRobustCommand{\VAN}[3]{##3}\VANthebibliography}

\defcitealias{Smith2022}{S22}
\newcommand{\stt}{\citetalias{Smith2022}}

\usepackage{graphicx}	
\usepackage{amsmath}	
\usepackage{longtable}

\title[Poisson-sampled stellar population models]
{Eigenspectra of Poisson-sampled stellar population models:\\Statistical fingerprints of cool giant stars in unresolved early-type galaxies}

\author[A. Gvozdenko and R. J. Smith]{
Anastasia Gvozdenko\thanks{E-mail: anastasia.gvozdenko@durham.ac.uk} and
Russell J. Smith
\\
Centre for Extragalactic Astronomy, Department of Physics, Durham University, South Road, Durham DH1 3LE, UK\\
}

\date{Accepted XXX. Received YYY; in original form ZZZ}

\pubyear{\the\year{}}

\begin{document}
\label{firstpage}
\pagerange{\pageref{firstpage}--\pageref{lastpage}}
\maketitle

\begin{abstract}

The light of unresolved galaxies carries the imprint of their underlying stellar populations, yet isolating contributions from specific stellar types remains challenging.
If multiple spectra of the same population are available, e.g. from high-resolution integral field unit (IFU) observations, random sampling variation can be exploited to extract information beyond that carried by the integrated light. 
We develop a principal component analysis (PCA) approach, applied to Poisson-sampled stellar population models, to probe the most luminous cool giant stars in old galaxies.
Across all wavelengths (0.45--2.4\,$\mu$m), and over all ages and metallicities explored, Poisson-sampling variation are dominated by a single PCA component. This leading component typically describes the most massive and coolest giant stars in the population. In the infrared, it resembles the integrated-light spectrum since the latter is already dominated by such stars; even so, all features from warmer stars are greatly suppressed. In the optical, by contrast, the method instead reveals cool M-giant contributions which are subdominant in the integrated light. 
A second PCA component is present in some bands, which is related to the variation in spectrum along the giant branch. 
Our model predictions are largely insensitive to age or the initial mass function, but show some dependence on metallicity, reflecting changes in the temperature and shape of the giant branch. The method can be applied to IFU observations in a purely data-driven way; comparison to our predictions will provide a new approach to testing some of the uncertain ingredients in stellar population models for early-type galaxies.

\end{abstract}




\section{Introduction}
The pre-eminent approach to probing the stellar content of galaxies too distant to resolve into individual stars, including the central regions of all massive early-type galaxies (ETGs), is through analysis of their integrated-light colours and spectra \citep{1980FCPh....5..287T,2013ARA&A..51..393C}. From a detailed comparison against evolutionary spectral synthesis models, this method seeks to infer the star-formation history, element abundance pattern, initial mass function (IMF), etc., which provide the best match to the observations. A wealth of detailed information on the stellar population trends in ETGs has emerged from this approach over many decades
\citep[e.g.][]{1998PASP..110..888W,2000AJ....119.1645T,2005ApJ...621..673T,2013MNRAS.429L..15F,2014ApJ...780...33C}.

Notwithstanding its widespread use, the integrated-light method nonetheless runs into some serious challenges. The total spectrum mixes the contribution of stars from all types of star (of all masses, in all evolutionary phases), leading to potential degeneracies and ambiguities in interpretation, especially where the effects on the spectrum are subtle. 
As an example, the inference of the IMF requires very precise knowledge of the cool giant\footnote{Throughout this work, we use the term \textit{cool giants} to refer to the cool, luminous post-main-sequence stars that contribute substantially to the integrated-light spectrum. For the ages considered here, this includes stars on the red-giant branch (RGB), red clump (RC), and asymptotic giant branch (AGB).} stars which dominate the integrated light in the red--optical. A small change to the assumed spectrum of these stars could be compensated
by altering the luminosity fraction attributed to cool dwarfs. In turn, a shift from 2\% to 5\% in this fraction amounts to the difference between Salpeter and Kroupa IMFs \citep[fig. 2 of][]{2012ApJ...747...69C}.
Unfortunately, the incidence and properties of cool giants are among the most uncertain ingredients in stellar population modelling. 
Notably, \cite{Conroy2012} found that their fits were very sensitive to including a `nuisance' parameter describing additional very cool giant contributions, beyond what was present in the model isochrones. Including this component generally reduced the need for bottom-heavy IMFs (their fig. 12).

There are also hints that current models still lack enough flexibility to match the spectra of very massive ellipticals. For example, the best current models systematically underpredict the strengths of the prominent CO bands in the near-infrared in massive ETGs \citep{Baldwin2018, Elham2022}. 
Recently, \cite{LaBarbera2025} investigated K-band absorption features in the massive elliptical galaxy M87 and concluded there must be contributions from a peculiar population of CO-strong, low-temperature stars which are absent from the models. It is unclear whether these stars should be dwarfs or giants. If the former, they would need to be very numerous to contribute enough light to the integrated spectrum, and hence directly related to the possibility of a bottom-heavy IMF in ETGs; if the latter, the implied mismatch with respect to model giants might still indirectly affect the IMF inference, for the reasons given above. 

Ultimately, the need for caution in interpreting ETG spectra stems from them having formed under conditions very different from the stars in the local neighbourhood from which our stellar population models are built and calibrated. 
In particular the cores of massive ETGs are expected to harbour stars born in intense rapid starbursts at early epochs, as witnessed in situ as dusty sub-millimetre galaxies \citep{Toft2014} or the even earlier systems being discovered with the {\it James Webb Space Telescope} (JWST) \citep{Cameron2023}.
Most notably, rapid star-formation events affect the abundance ratios, and much effort has been put into accounting for $\alpha$- and Na-enhancements, and more detailed abundance patterns, and their effect on the spectra through differential corrections applied to empirical stellar libraries \citep[e.g.][]{2012ApJ...747...69C,2017MNRAS.464.3597L,2023MNRAS.523.3450K}.

Even more exotic deviations from the Milky Way disk conditions could also be at play. For example, the ultraviolet excess seen in the most massive ETGs \citep{1999ARA&A..37..603O}  requires very hot evolved stars which are difficult to reproduce in canonical models at high metallicity. One hypothesis for their formation appeals to enrichment in helium, for at least part of the population \citep{Chung2017,Ali2023}. Because main-sequence lifetimes are dramatically shorter for higher helium fraction, this leads to lower-mass stars in all evolutionary phases at given age, e.g. RGB stars having $\sim$0.6\,$M_{\odot}$. Although potentially `solving' the ultraviolet-excess problem (because lower-mass stars reach higher temperatures on the horizontal branch), this proposal would open the question of whether the evolutionary tracks of such stars are sufficiently well known for the cooler phases driving the optical and infrared spectra. The possibility that such stars might skip the asymptotic giant branch (AGB) altogether \citep{1990ApJ...364...35G,2015A&A...578A.117C,Costa2024} further complicates the picture. 

Given these challenges, it is desirable to explore alternative methods beyond the standard integrated-light technique to probe the stellar content in unresolvable populations. 
One promising avenue lies in `variational' techniques, analysing the differences among independent random samples of stars from a given stellar population, e.g. as probed by the individual pixels (or resolution elements) of an image or IFU datacube.  
In images, the most luminous stars, typically RGB and AGB stars, make the dominant contribution to measurable surface-brightness fluctuations (SBFs)  \citep{Tonry1988}. The ratio of variance to mean surface brightness can be interpreted as a flux heavily weighted to that of the brightest individual stars present, which can be used not only as a distance indicator, but also as a probe of the evolved stellar populations \citep[e.g.][]{1993ApJ...409..530W,2007ApJ...668..130C,2021MNRAS.507.3005R}.
The stochastic variance in an integrated stellar population arises from the underlying distribution of stellar luminosities, with Poisson sampling providing one way of generating stochastic realisations of the population (e.g. \citealt{Buzzoni2005}).

Previous studies have explored the stellar populations responsible for SBFs and how their contributions may be considered across different wavelength ranges \citep{Buzzoni1989, 1993A&A...275..433B,2020MNRAS.493.5131V}. Extending this approach to the spectral domain allows us to investigate how stellar contributions to integrated-light spectra vary with wavelength.
The most straightforward approach to this is to apply the standard imaging SBF measurement to each slice of an IFU datacube, treating the wavelength channels completely independently. \cite{Mitzkus2018} presented the first observational use of this method, which in principle measures the flux of the brightest stars contributing at each wavelength (which can be different types of stars at each wavelength).
An alternative approach was developed by \cite{vanDokkum2014}, who calculated the expected ratio of spectra between high- and low-SBF pixels in a population model (the SBF defined according to broadband flux). This `fluctuation spectrum' differs from the channel-wise SBF because it captures the inherent correlation in the data, i.e. that brighter pixels are those specifically with more cool giant stars, and hence the features of these stars are extracted, at all wavelengths. They applied this method only to observed narrow-band colours, rather than to spectra.

\citet[][hereafter \stt]{Smith2022} proposed an alternative approach based on principal component analysis (PCA) applied directly to unresolved spectra from individual spatial pixels of an IFU datacube. By decomposing the pixel-to-pixel spectral variations into a small number of orthogonal components, or `eigenspectra', this method captures correlated structure in the dataset, driven by the presence and properties of the most luminous giant stars. In this respect, the approach is more similar to that of \cite{vanDokkum2014} than to \cite{Mitzkus2018}, but it is purely spectroscopic without reference to the pixel fluxes, and is more general, being able to extract multiple independent fluctuation signals if present.
\stt\ demonstrated the characteristic form of these eigenspectra in the I-band for a single 10\,Gyr solar-metallicity stellar population, and showed that the expected signal can be recovered in practice from high-angular-resolution MUSE observations. Notably, the analysis was able to isolate the spectrum of the coolest giant stars, which (in models) collectively contribute only $\lesssim$10 per cent of the total integrated light.

In this paper, we further develop the spectral fluctuations method, seeking (1) to quantify how the expected signals vary with stellar population parameters, beyond the single age and metallicity considered in \stt\ and (2) to explore the information content of the eigenspectra over a broader wavelength range spanning the optical and near-IR regime. These are the necessary first steps towards optimising strategies for observations with current and future facilities such as ERIS \cite{2023A&A...674A.207D} and MAVIS \citep{2020arXiv200909242M} at the ESO VLT, and HARMONI on the ESO ELT \citep{Thatte2022}.

In Section~\ref{sec:meth}, we review the principles of the method and describe the components of the models employed, illustrating the approach through a reanalysis of the I-band, 10\,Gyr, solar-metallicity stellar population case for comparison with \stt. Section~\ref{sec:wavs} then presents the extension to different wavelength ranges, while Section~\ref{sec:pars} explores a wider range in stellar population parameters. In Section~\ref{sec:disc} we briefly discuss some of the strengths and current limitations of our methods, as well as addressing the observational applicability of the method and its practical requirements. 
Our conclusions are summarised in Section~\ref{sec:concs}.

\section{Methods}
\label{sec:meth}

In this section, we first describe the generation of  Poisson-sampled realisations from stellar population models (Section~\ref{sec:poisamp}), and then the treatment of these through the PCA approach (Section~\ref{sec:pca}), before presenting in detail the results for the particular case previously analysed in \stt\ as an illustrative example (Section~\ref{sec:base}).

\subsection{Poisson-sampled simple stellar populations}\label{sec:poisamp}

Classical population-synthesis models assume a well-populated stellar population, with the number of stars along the isochrone determined by the underlying IMF \citep{Tinsley1980, Buzzoni1989}.
We implement relatively simple spectral synthesis models which are intended to capture, in particular, the information carried by Poisson sampling of the cool giant stars. 
We neglect variation of abundance ratios, and restrict our modelling to the stellar population ages greater than 3\,Gyr for this study, in order to avoid uncertain model treatment of thermally-pulsing AGB stars, or the possible presence of carbon stars. Nonetheless, our treatment spans a range of older ages (3, 6, 10, and 13~Gyr) and metallicities (-- 0.3, 0.0, and + 0.3~dex) to explore whether the results are sensitive to these parameters. 
Here, we generate a large set of stochastically sampled stellar populations, following the approach of \stt, to investigate how Poisson sampling of cool giant stars affects integrated-light spectra and the resulting surface-brightness fluctuations (SBFs). By retaining information about the stellar populations in individual isochrone bins, this modelling framework is not limited to predicting SBFs, but can also be used to investigate other properties and correlated variations in integrated-light spectra, as discussed in the Introduction.

The spectra are synthesised following a similar procedure to that described in \cite{Smith2022}, to maintain consistency with the earlier analysis. 
We adopt published isochrones from the PAdova and TRieste Stellar Evolution Code (PARSEC; version 1.2S; \citealt{Bressan2012}) with the COLIBRI extension for the AGB (S\_37; \citealt{Pastorelli2020}). For each bin provided in the isochrone, a stellar spectrum is generated using the
{\sc spigen}\footnote{https://github.com/AlexaVillaume/SPI\_Utils} polynomial interpolator, which employs empirical spectra from MILES \citep{Sánchez-Blázquez2006} and Infrared Telescope Facility library (IRFT; \citealt{Villaume2017}), together covering a wavelength range 0.35--2.5$\mu$m. The stellar parameters associated with each point in the isochrone table are used to select the corresponding stellar spectrum. 

To populate the isochrone, we adopt a single power-law IMF, $\mathrm{d}N/\mathrm{d}M \propto M^{-\alpha}$, with fixed exponent $\alpha=2.35$ \citep{Salpeter1955}. Because the spectral fluctuations are driven by stars in a very narrow range in mass, there is no need to explore different IMF treatments. 
We use the IMF as a weighting function to determine the relative expected number of stars in each isochrone bin, by integrating the IMF over the corresponding initial-mass interval. For each mass interval, the stellar atmosphere parameters are taken from the lower mass of the two bounding tabulated isochrone points. Consequently, the lowest-mass point of the isochrone is not assigned a separate atmosphere model, as it does not define an additional finite mass interval for the IMF integration, but the contribution of these stars is anyway negligible. The relative weights are then normalised to obtain the mean absolute number of stars by scaling the population to a given assumed total luminosity, which corresponds to the luminosity, e.g. of one pixel in an IFU observation. 
For each set of bandpass, age and metallicity, we adopt a total luminosity of 10\,000\,$L_\odot$ in the passband considered.\footnote{The value itself is arbitrary, and all of our results are independent of this choice, so long as all necessary stars are sampled {\it on average}, over the whole dataset. For reference, 10\,000\,$L_\odot$ corresponds to the luminosity sampled by a 0.05--0.10\,arcsec pixel-scale at around the effective radius, in an elliptical galaxy at 10\,Mpc, depending on bandpass.} Since the luminosity contributed by individual stars depends on the passband, the total number of stars in the simulations also differs between passbands, resulting in different sampling statistics. Thus, simulations in different passbands represent independently sampled stellar populations rather than different wavelength ranges of the same simulation. 
Finally, the integer number of stars in each isochrone bin is drawn from a Poisson distribution with the corresponding mean number of stars, to generate a stochastic `realisation' of the population. 
Given the fine mass-sampling of the isochrones for the evolved stars, there are often some isochrone bins which  are assigned no stars at all in an individual realisation. However, over a sufficiently large number of independent realisations, the mean number of stars in each bin converges to the expected value.

To focus on narrow-band features and to match realistic observational use-cases, the synthesised spectra are divided into individual windows (here referred to as V, I, J, H, and K, although the correspondence with photometric bands is only approximate). 
The final synthesised spectrum for each realisation is constructed by first normalising each {\sc spigen} stellar spectrum to unit transmission-weighted flux using the corresponding filter transmission profile from the SVO Filter Profile Service\footnote{http://svo2.cab.inta-csic.es/svo/theory/fps3/index.php} for the photometric band considered (V, I, J, H, or K). Each spectrum is then rescaled to the luminosity of the corresponding isochrone point in that band, derived from the photometric magnitude tabulated in the PARSEC+COLIBRI isochrone using the corresponding solar absolute magnitude. The resulting stellar spectra are multiplied by the Poisson-sampled number of stars in each isochrone mass interval and summed to obtain the integrated spectrum of each realisation.
This treatment does not explicitly distinguish between different evolutionary stages at the same stellar parameters. In particular, RGB and AGB stars with similar atmospheric parameters are assigned empirical spectra from SPIGEN selected according to those parameters, irrespective of their evolutionary stage. 

Our results are typically based on 1000 realisations of each model, with each realisation representing an independent sampling of a stellar population
restricted to the wavelength range of a given passband.
This number of realisations is sufficient to ensure stability in the derived eigenspectra. Since the realisations are wholly independent from each other, they correspond to observed `pixels' in an IFU cube, only after binning on a sufficient scale to remove PSF effects.

\subsection{Principal Components Analysis}\label{sec:pca}
\label{sec:pca_theory}

PCA is a long-established technique to reduce the dimensionality of a multivariate dataset. The method was introduced by \citealt{pearson1901} and has been applied very widely \citep[e.g.][]{jolliffe2016, Beattie2021}. Some notable applications in astronomical spectroscopy include those of \cite{1992ApJ...398..476F} for quasars,  \cite{1995AJ....110.1071C} for galaxies, and \cite{1998MNRAS.298..361B} for stars.  PCA simplifies complex data by transforming the original variables into a smaller set of linearly independent components, while retaining the largest possible fraction of the variance present in the data. 
As a purely data-driven method, PCA requires no prior knowledge of the physical origin of the patterns in the observations, nor any auxiliary information beyond the dataset itself. 
However, to interpret real observations correctly, it is often helpful to apply the same analysis to mock data or to noise-free models \citep[e.g.][]{2007MNRAS.381..543W}. In our analysis, the `data' comprise the set of Poisson realisations from each stellar population model, for the particular band being studied. The spectra are used at their full native resolution to retain the full information content in the PCA, though we do not expect that modest spectral smoothing would greatly affect the results.
In preparation for the analysis, we independently fit a first-order polynomial to each integrated-light realisation over the complete wavelength interval considered for each band and divide the spectrum by the fitted polynomial. This removes the overall flux normalisation and broad spectral slope, reducing the sensitivity of the PCA to broad-band colour and flux-calibration differences that may affect practical IFU observations.

The principal components (PCs) are ordered according to their contribution to the total variance in the dataset. The first few PCs typically capture the majority of the information while excluding much of the noise (particularly relevant for observational data). 
In the case of simulated data, the leading components similarly describe the dominant sources of variance, with the later PCs representing incoherent or statistically insignificant features that can often be neglected. 
For this paper, we write the eigenspectrum of the $i$'th PC as $E_i$, with its corresponding eigenvalue for the $j$'th input spectrum being $P_{i,j}$. (For brevity, the set of all $P_{i,j}$ values is just $P_i$.) Then the $j$'th spectrum is given by
\[
S_j(\lambda) = \sum_{i} P_{i,j} \,  E_i(\lambda) \ .
\]

The present study focuses mainly on the eigenspectra, since it is these which carry the physical information about the luminous cool giant stars; the eigenvalues merely reflect the random sampling of stars in each realisation of the model. Given that PCA is a mathematical operation with no knowledge of the underlying physics, each resulting component can appear with an arbitrary sign. For clarity, we present the eigenspectra with the sign chosen so that the strongest or most familiar spectral features appear in absorption where expected, at least for the first component. 

It is important to note that when PCA is applied to spectra, the resulting components are well-defined only for the {\it whole} wavelength interval over which the analysis is performed, and not for each specific wavelength channel individually. 
Hence, it will be particularly important to match the wavelength ranges carefully to perform a fair comparison, e.g. between models and observed spectra.

\subsection{The reference case: I-band for 10~Gyr, solar metallicity}
\label{sec:base}

\begin{figure*}
\centering
\includegraphics[width=17cm]{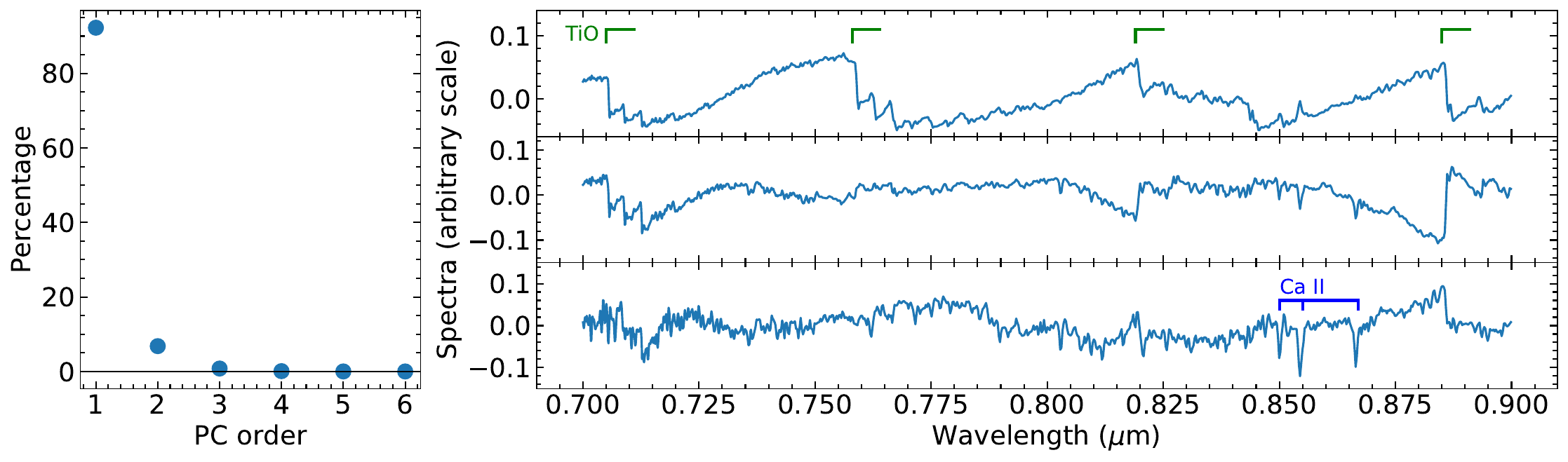}
\caption{PCA results for the 10 Gyr, solar metallicity reference case. Left: the fractional contributions to the total variance of the first six PCs.
Right: the first three eigenspectra of the spectra in I-band, with prominent spectral features indicated.
 }
 \label{fig:Base-case-ESR}
\end{figure*}

\begin{figure*} 
 \includegraphics[width=18cm]{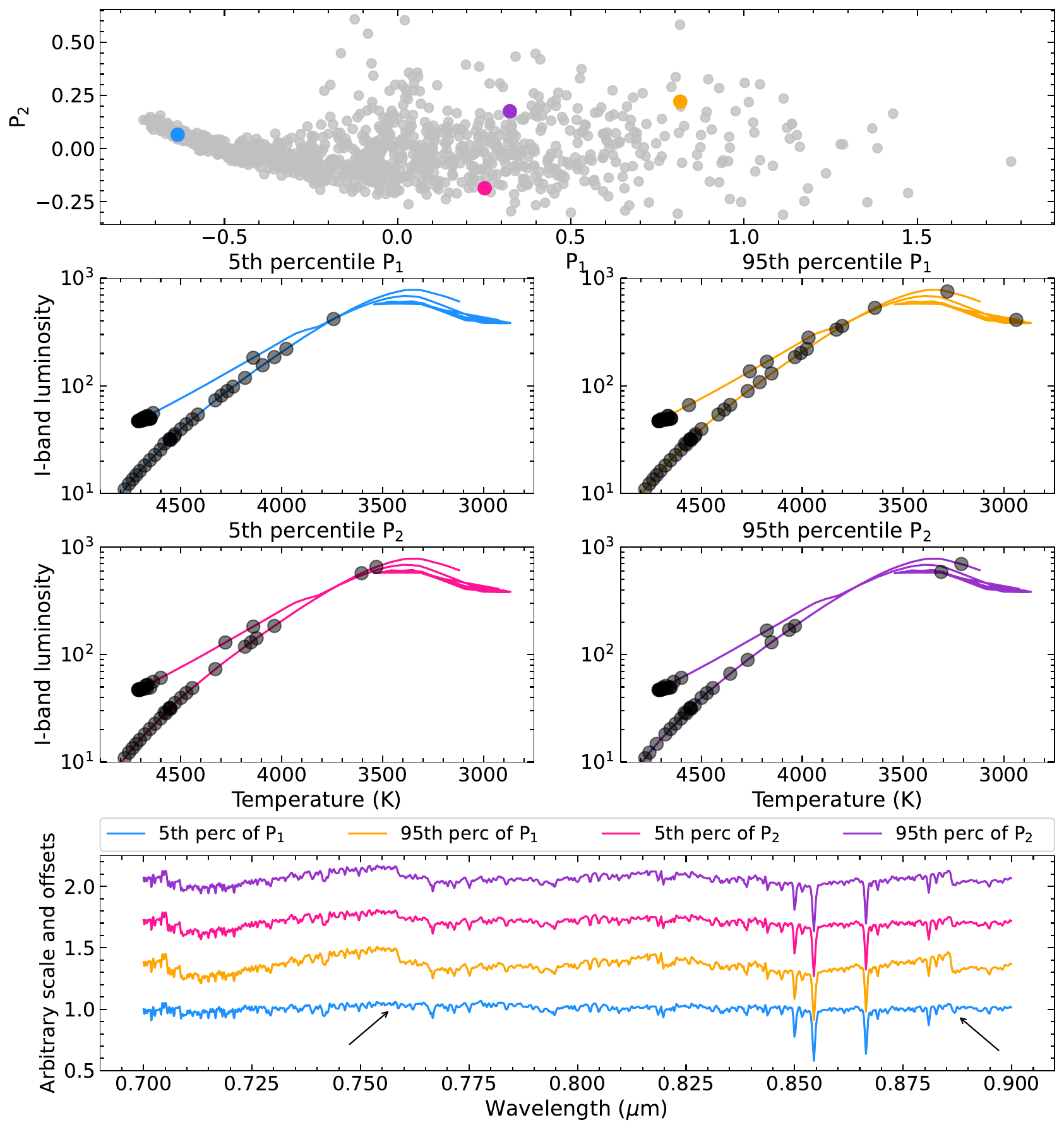}
 \caption{
 Eigenvalue space plot for solar metallicity and 10\,Gyr isochrone case in the I-band. Top: second eigenvalues versus the first eigenvalues; grey points represent all of the realisations, while coloured points are illustrative selected realisations (blue: 5th percentile of $P_1$; orange: 95th percentile of $P_1$; pink: 5th percentile of $P_2$; purple: 95th percentile of $P_2$).
 Middle: the isochrone paths of corresponding colours for each of the realisations shown in band luminosity versus temperature. The dark circles indicate the populated isochrone bins. Bottom: the synthetic spectra of the four illustrative realisations, with the same colour coding as above. The arrows highlight some regions in the spectra where the differences are visually most noticeable.}
 \label{fig:main-Iband-t10p00m}
\end{figure*}

\begin{figure*}
 \includegraphics[width=17cm]{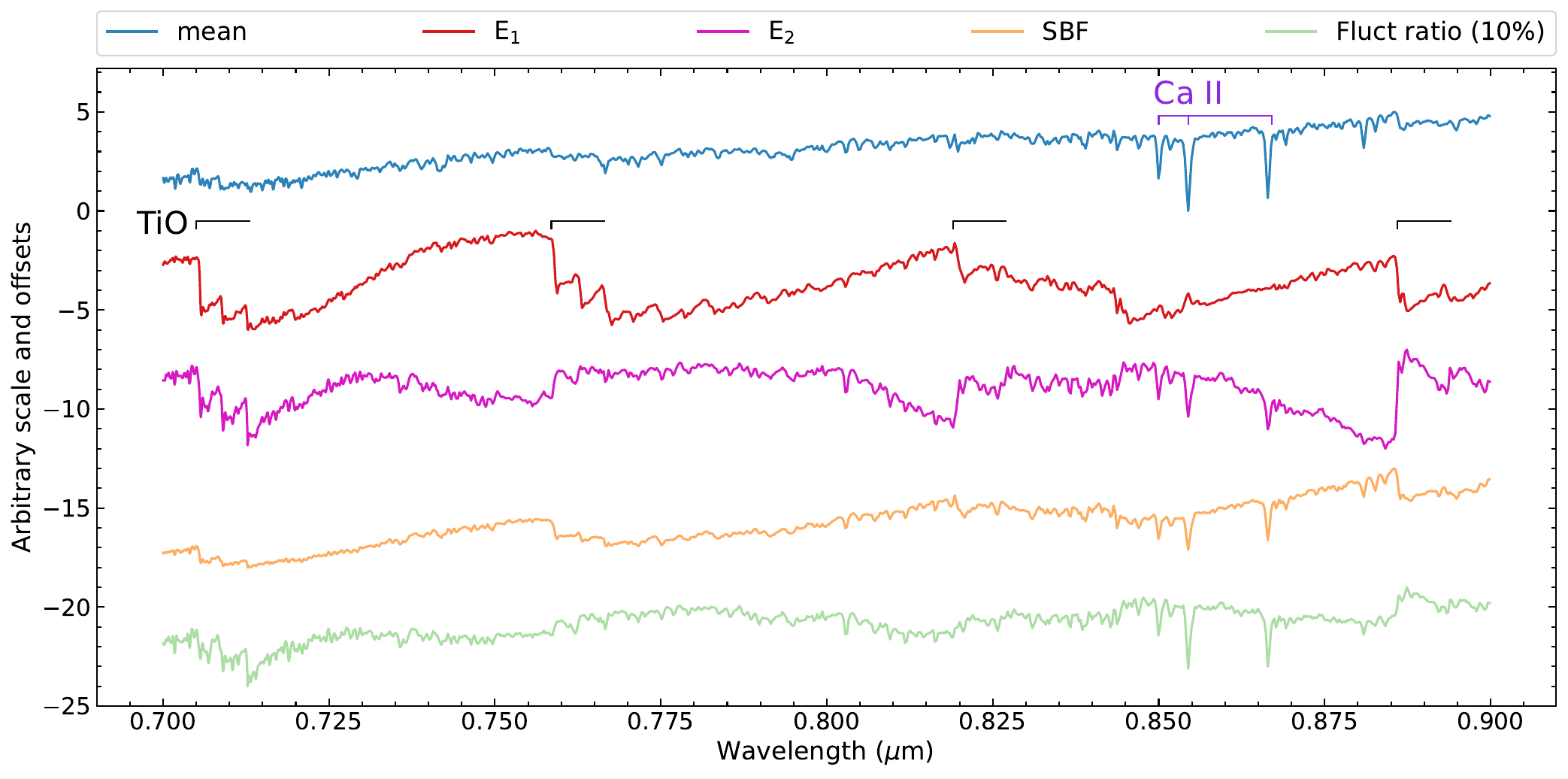}
 \caption{Eigenspectra, SBF, and fluctuation spectroscopy methods applied to I-band wavelength of solar metallicity and 10 Gyr isochrone. From top to bottom: the blue line shows the mean spectrum (not normalised), the red line is the first eigenspectrum, the pink line is the second eigenspectrum, the orange line is the spectrum obtained with the SBF method \protect\cite{Mitzkus2018} definition, and the green line is the fluctuation ratio spectrum from \protect\cite{vanDokkum2014} definition, with some prominent spectral features indicated. }
 \label{fig:spectral_plot-Iband-t10p00m}
\end{figure*}

\stt\ explored the case of a 10~Gyr solar-metallicity stellar population model, in the I-band. We begin by presenting results in detail for this same configuration as a reference case (Figures~\ref{fig:Base-case-ESR}--\ref{fig:spectral_plot-Iband-t10p00m}), to introduce some terminology and diagnostic figures that will be used throughout the paper. 
In comparing the results here to those in \stt, we note that the key differences in treatment are (1) the inclusion of the COLIBRI-extended tracks in the isochrones used to synthesise the spectra, and (2) no spectral smoothing has been applied to the input data. 

The left panel of Figure\,\ref{fig:Base-case-ESR} shows the fractional variance captured by each of the first PCs. The data that were passed to the PCA has no noise, and therefore, this result shows the information content of the models. The addition of noise would introduce a `floor' to the variance contributions, preventing the information in later PCs from being recovered (see fig,~9 of \stt).
Figure\,\ref{fig:Base-case-ESR} shows that the overwhelming bulk of the information resides within the first two eigenspectra (99 per cent of variance in this case). 
This dominance of the first two components holds across all cases examined in this paper. Accordingly, our subsequent analyses focus exclusively on the first two PCs. 

Turning to the eigenspectra themselves, we confirm the result of \stt\ that the leading component has the form of a late-M type giant, essentially the spectrum of stars at the tip of the RGB/AGB in this population. The spectrum is dominated by the Ti\,O ($\gamma$- and $\phi$-system) band-heads. The second eigenspectrum has the form of a wavelength-dependent `correction' to these bands: adding more of $E_2$ strengthens the absorption features at shorter wavelengths, and weakens those towards the red. The third component shows more narrow atomic features, notably the Ca\,{\sc ii} triplet lines, though some features from Ti\,O are still noticeable. Again, these results are all qualitatively similar to those of \stt.

The eigenvalues of each realisation correspond to the relative weights of the components, and depend upon the particular set of stars contributing to the spectrum. 
Figure\,\ref{fig:main-Iband-t10p00m} explores the relationship between eigenvalues, isochrone sampling and spectral realisations. 
We highlight four particular realisations which differ substantially in the contribution of the first two components, as quantified by the eigenvalues $P_1$ and $P_2$ (upper panel).
For each of these realisations, the middle set of panels shows how the isochrone was populated by the Poisson-sampled model.
This representation demonstrates that the first eigenvalue ($P_1$) traces the presence of the most luminous stars: higher values of $P_1$ correspond to realisations containing a few extremely bright giants near the tip of the RGB, while lower $P_1$ values represent realisations where the light is dominated by a larger number of fainter stars, with no bright giants present. 
The correspondence of $P_2$ with the isochrone population is a little less obvious, but (as also noted in \stt) this component is related to the  `evenness' of sampling along the giant branch. A realisation with high value of $P_2$ but similar $P_1$ typically has its giant light concentrated right at the tip of the branch, often with a `gap' in the distribution at lower masses. 
Finally, the lower panel of the figure shows the synthesised spectra for the same four realisations of the model. The differences caused by the range in $P_1$  is evident, with low values corresponding to `flatter' spectra, while the breaks and troughs caused by Ti\,O bands being visually more prominent for high $P_1$, as highlighted.
As may be expected, the effect of the second component is much less striking. Guided by the form of $E_2$ though, it is possible to confirm that the high-$P_2$ realisation has a stronger 0.885\,$\mu$m Ti\,O band than the low-$P_2$ spectrum, though their 0.705\,$\mu$m bands are similar. 

Figure\,\ref{fig:spectral_plot-Iband-t10p00m} compares the eigenspectra from the PCA approach against the ordinary integrated light spectrum, and against two alternative techniques for analysing the Poisson-sampling variation in spectra: the per-channel SBF spectrum, following the prescription of \cite{Mitzkus2018} and the `fluctuation ratio spectrum' following the approach of \cite{vanDokkum2014}\footnote{The fluctuation ratio spectrum is defined here as the mean spectrum of the pixels in the brightest decile, divided by the mean spectrum of the faintest decile, in the relevant band.}.
For the 10\,Gyr solar-metallicity, I-band case, the comparison confirms that all three of these `fluctuation' methods
are selectively sensitive to the cool giant stars, with the Ti\,O bands emphasised, as compared to the integrated light. 
There are, however, some notable differences between the information picked up by the three methods, e.g. with the Ca\,{\sc ii} lines being more prominent in the SBF spectrum than in $E_1$ or the ratio. (This aspect of the variation  then appears to be picked up in $E_2$ instead, our method.)
We comment further on these differences below, especially regarding the V-band, where they are most pronounced.

In the I-band, for the old, solar-metallicity population, the incorporation of COLIBRI into the isochrones appears to slightly alter the second eigenspectrum, as shown in
Appendix\,\ref{apx:colibri}, whereas the first eigenspectrum, which carries the great majority of the extracted variance, remains largely unaffected. Moreover, the new isochrones and the addition of COLIBRI result in a substantially smaller relative variance contribution from the second PC (and greater contribution from the first) compared with the fractions reported by \stt. 
At such old ages, the extension to the thermally pulsing AGB phase contributes only $\sim$0.5\ per cent to the I-band output in these models; differences in the PCA results probably arise instead from the early AGB ($\sim$9\ per cent), which is slightly cooler and fainter in COLIBRI than in the PARSEC isochrone.


\section{Exploration of wavelength ranges}
\label{sec:wavs}

In this section, we expand our discussion of the 10\,Gyr solar-metallicity case, to include wavelength intervals beyond the I band. 

The synthetic spectra generated by {\sc spigen} span 0.36–2.50 $\mu$m. We divide this range into five regions, which we (loosely) refer to as the V, I, J, H, and K bands.
Dividing the full spectrum into separate wavelength bands reflects likely practical use cases and allows us to identify the distinct stellar contributors that drive the fluctuation signals in different regions of the spectrum. This treatment also avoids impact from larger-scale trends, such as the low-order continuum shape, which could potentially obscure the narrower feature variations that our method is designed to detect.

Although the mean total luminosity is held constant in solar units across all bands, the fluctuation amplitude increases markedly towards the red, as expected from the wavelength-dependent contribution of the most luminous stars to the integrated light (e.g. \citealt{Buzzoni1989, Cervino2006}). 
The effective number of stars, $\bar{N}$, reflects the luminosity-weighted contribution of stars to the integrated light rather than the actual number of stars in the population or in a particular evolutionary phase (e.g. N$_\text{ev}$ for post-MS stars ).
This wavelength dependence arises from the changing luminosity weighting of the stellar populations: while luminous cool red-giant stars contribute substantially to the V-band luminosity, they increasingly dominate the luminosity towards the red, where a smaller effective number of stars contributes to the integrated light. The resulting decrease in $\bar{N}$ leads to larger stochastic fluctuations at longer wavelengths.
Noticeably the provided values of the total number of stars in post-MS evolutionary stages decrease from V to K similarly to $\bar{N}$, however the order of this values do not always agree.  
Table\,\ref{tab:basecases-stat} lists, for each wavelength band, the fractional variance contributions of the first four components in the PCA decomposition, together with the quantities $\bar{L} = {\rm var}(L)/\langle{}L\rangle$, which is the classic SBF luminosity (i.e. characteristic luminosity of the stars driving the fluctuations), and also $\bar{N} =  \langle{}L\rangle/\bar{L} = \langle{}L\rangle^2/{\rm var}(L)$, which is the effective number of stars per sample \citep{1993A&A...275..433B}.
As expected, the fluctuations in the redder bands are dominated by only a handful of very bright red giants per synthesised spectrum.
In all bands, the variation among samples is dominated by the leading components, with $>95$ per cent of the variance accounted for by PC1 and PC2. (See Appendix\,\ref{App-table} for equivalent results for the full set of cases explored.)

\begin{table*}
   \caption{The statistical characteristics for each base case, solar 10 Gyr, in each of the bands explored. The percentage carried by the first four PCs for each wavelength band is highlighted in bold if it is greater than 5\%. The last two columns are the SBF luminosity $\bar{L}$ (the characteristic luminosity of the stars driving the Poisson fluctuations), and the $\bar{N}$, the effective number of such stars per computed sample. By definition, the product $\bar{N}\bar{L} = \langle{}L\rangle \approx 10\,000\,L_\odot$ in all cases, this being the mean luminosity per sample from the population. The number of evolved stars from the isochrone that include stars from RGB to post-AGB is given by N$_\text{ev}$.}
    \label{tab:basecases-stat}
    \centering
\begin{tabular}{ccrrrrccc}
Band  & $\lambda$ (nm) & PC1 & PC2 & PC3 & PC4  &   
$\bar{L}/L_\odot$ & 
$\bar{N}$ &
N$_\text{ev}$
\\
\hline

V & 0.45--0.70 & \bf{89.7} & \bf{9.0} & 0.8 & 0.4 & \phantom{00}28 & 360\phantom{.0}   & 250\\
I & 0.70--0.90 & \bf{92.2} & \bf{6.8} & 0.8 & 0.1 & \phantom{0}130  & \phantom{0}78\phantom{.0} & 160\\
J & 0.90--1.30 & \bf{78.8} & \bf{17.6} & 2.8 & 0.7  & \phantom{0}900 & \phantom{0}11\phantom{.0} & 95\\
H & 1.40--1.85 & \bf{92.6} & \bf{6.3} & 1.0 & 0.1 & 1800 & \phantom{00}5.5  & 67\\
K & 1.95--2.40 & \bf{95.6} & 4.2 & 0.2 & $<$0.1  & 2800  & \phantom{00}3.6 & 59\\

\end{tabular}
\end{table*}

Having already described the I-band results, we now proceed to longer wavelengths, returning finally to the V-band where some contrasting behaviour is observed.

\begin{figure*}
 \includegraphics[width=17cm]{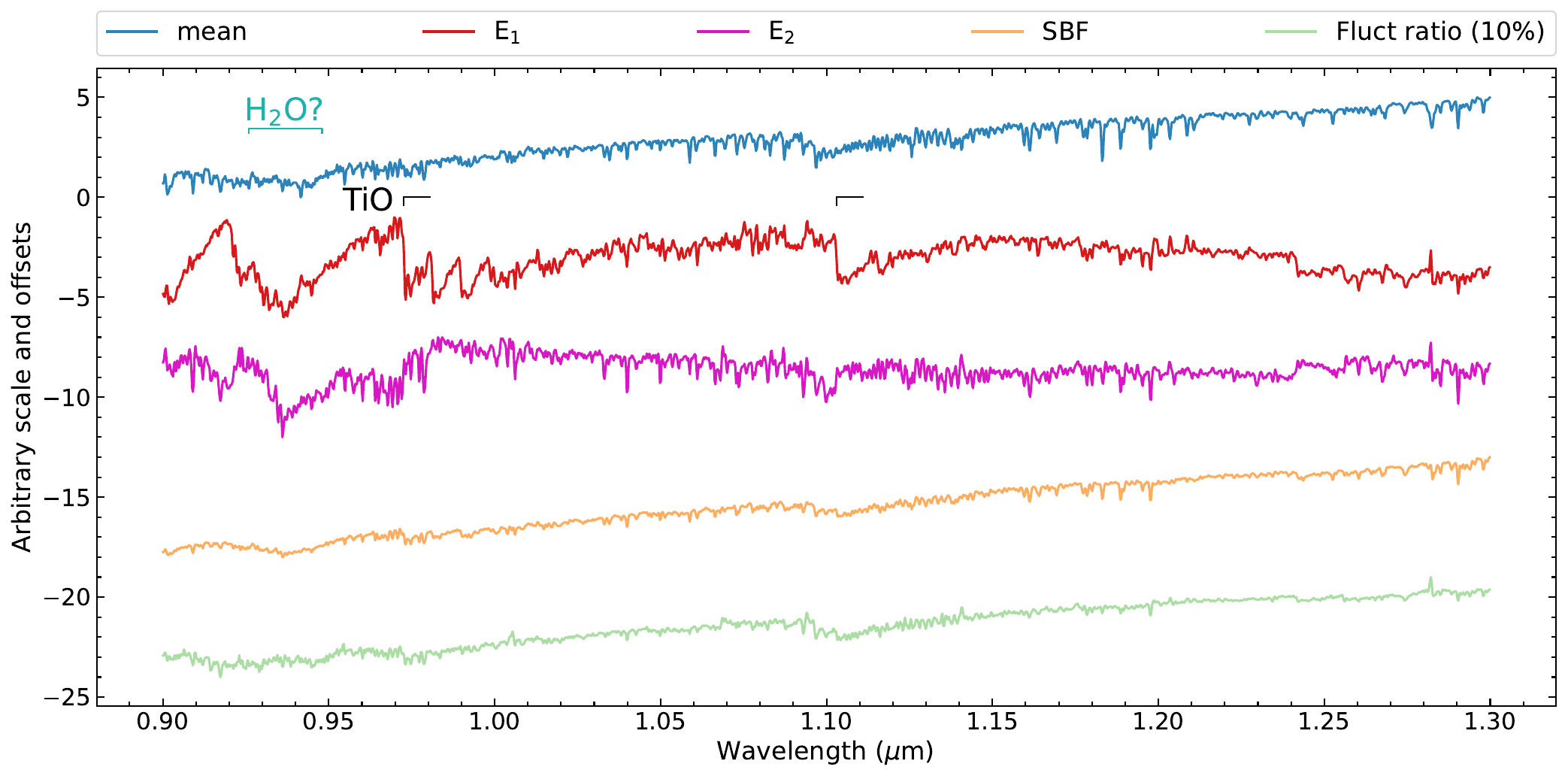}
 \caption{Same methods as in Figure\,\ref{fig:spectral_plot-Iband-t10p00m}, applied to the J-band wavelength of solar metallicity and 10 Gyr isochrone.}
 \label{fig:spectral_plot-Jband-t10p00m}
\end{figure*}

The results for the J band (here defined as 0.9--1.3\,$\mu$m, i.e. including parts of the z and Y bands)  are shown in Figure~\ref{fig:spectral_plot-Jband-t10p00m}.
The J-band integrated-light spectrum is characterised by fairly weak atomic absorption features for old populations. The broad feature near $\sim$1.1\,$\mu$m blends absorption bands from  H$_2$O, Ti\,O and CN \citep{Joyce1998,Lancon2000}. In the first eigenspectrum (and in the fluctuation ratio spectrum), a sharp break at 1.104\,$\mu$m can be best identified as the Ti\,O $\phi(0-0)$ head.
The strong feature seen at 0.935\,$\mu$m in $E_1$ is present in the stellar spectra and must be a trait of cool giants. A similar band was identified by \cite{vanDokkum2014} as H$_2$O, but \cite{Lancon2000} identify a Ti\,O $\epsilon(\Delta{}v\,=\,-1)$ band here in the spectra of cool long-period variables. The `sawtooth' of bandheads at 0.97--1.00\,$\mu$m is the Ti\,O $\delta(\Delta{}v\,=\,-1)$ sequence. 
Among all of the wavelength ranges explored here, the J band has the smallest contribution of variance from the first PC, and the largest contribution from the second component. The additional information carried by $E_2$ is largely concentrated in the $<$1\,$\mu$m region, effectively describing a balance between the Ti\,O and (possible) H$_2$O features.

\begin{figure*}
 \includegraphics[width=17cm]{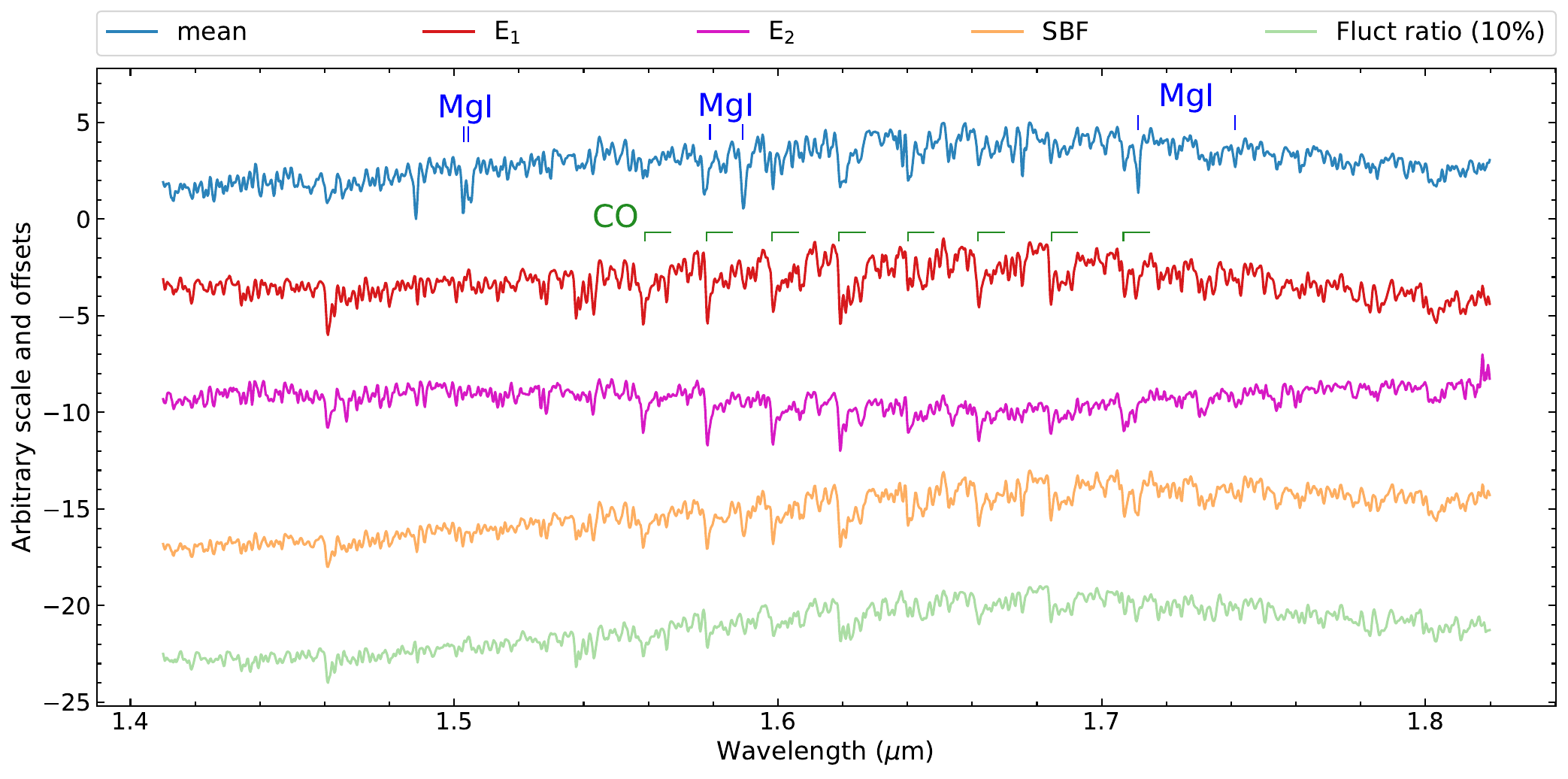}
 \caption{Same methods as in Figure\,\ref{fig:spectral_plot-Iband-t10p00m}, applied to the H-band wavelength of solar metallicity and 10 Gyr isochrone.}
 \label{fig:spectral_plot-Hband-t10p00m}
\end{figure*}

In the H band, the cool giants dominate the integrated light, and hence the distinction between $E_1$ and the mean spectrum becomes less pronounced. 
The H-band spectra are characterised by a curved continuum, which is not removed by our linear normalisation, and it is notable that $E_1$ and $E_2$ seem to pick up some variation in broad-band shape between the sample spectra, rather than adding distinct information about the narrower absorption features. The variation in curvature might arise from differences in the sampling of very late M giants, which have strong H$_2$O absorption at towards both sides of the wavelength window.
The atomic features (notably Mg\,{\sc i} lines) are suppressed in $E_1$, leaving only the second-overtone CO bandheads through most of the spectral range. At the short end, a narrow feature of unknown origin appears at $\sim$1.46\,$\mu$m\footnote{This wavelength is in a region of strong atmospheric absorption, so the stellar library spectra from which the models are built might be less reliable here.}. The H-band SBF spectrum and fluctuation ratio spectrum are qualitatively similar to $E_1$ (see Figure\,\ref{fig:spectral_plot-Hband-t10p00m}).

\begin{figure*}
 \includegraphics[width=17cm]{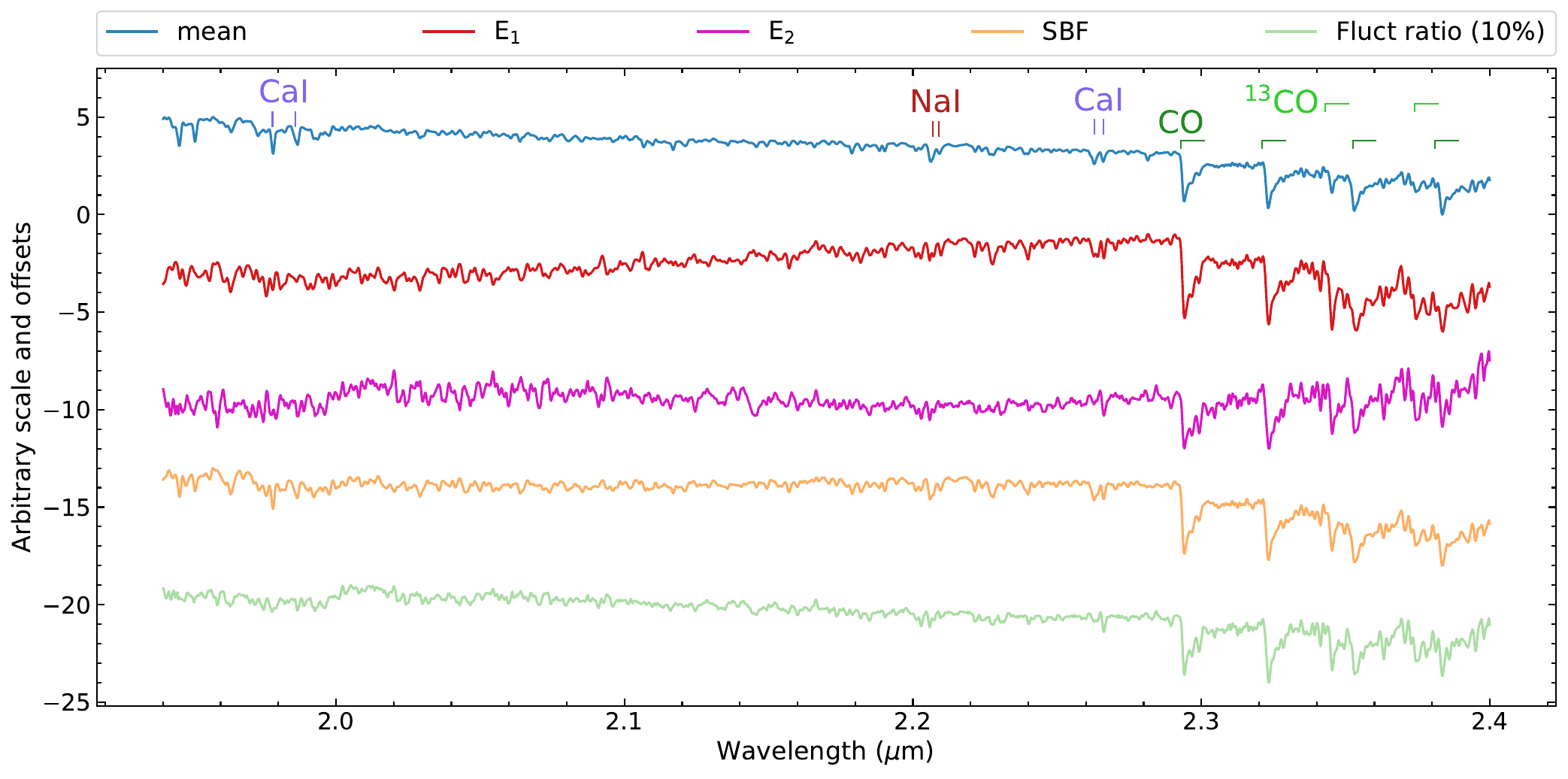}
 \caption{Same methods as in Figure\,\ref{fig:spectral_plot-Iband-t10p00m}, applied to the K-band wavelength of solar metallicity and 10 Gyr isochrone.}
\label{fig:spectral_plot-Kband-t10p00m} 
\end{figure*}

Similarly to H, the K-band integrated light is already dominated by cool giant stars, hence the differences between $E_1$ and the mean spectrum are slight. The few atomic in Figure\,\ref{fig:spectral_plot-Kband-t10p00m} features (mainly Ca\,{\sc i} and Na\,{\sc i})  are suppressed in $E_1$, though the 2.26\,$\mu$m Ca\,{\sc i} doublet persists as it originates in giants. The red end of the spectrum is dominated by the CO first-overtone bandheads, carried by giants, and hence persists in the fluctuation methods. 
Suggestively, the $^{13}$CO bandhead is strengthened in $E_1$ compared to the integrated light, consistent with the surface abundance of $^{13}$C being enhanced by dredge-up and/or extra mixing in the more evolved stars driving the fluctuations \citep{Gilroy1991}\footnote{Our models are not ``predictive'' on this point, however, since the stellar library does not distinguish AGB vs RGB stars, but further work could explore the effect.}.
Again, $E_2$ appears to account for slight continuum shape differences, rather than narrow features. 

\begin{figure*}
 \includegraphics[width=18cm]{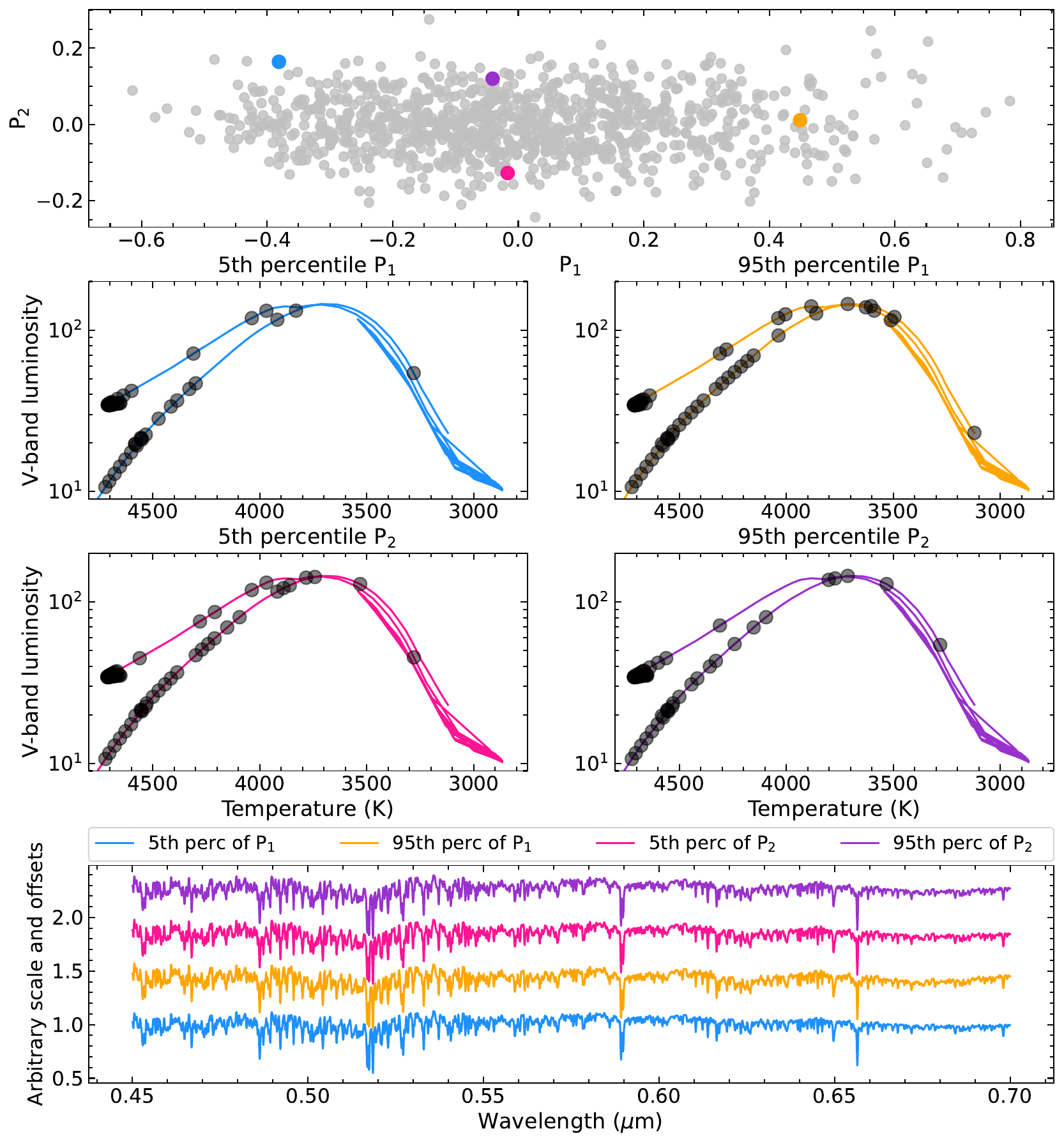}
 \caption{Same as Figure\,\ref{fig:main-Iband-t10p00m} for solar metallicity and 10\,Gyr isochrone case in the V-band.}
 \label{fig:main-Vband-t10p00m}
\end{figure*}

\begin{figure*}
 \includegraphics[width=17cm]{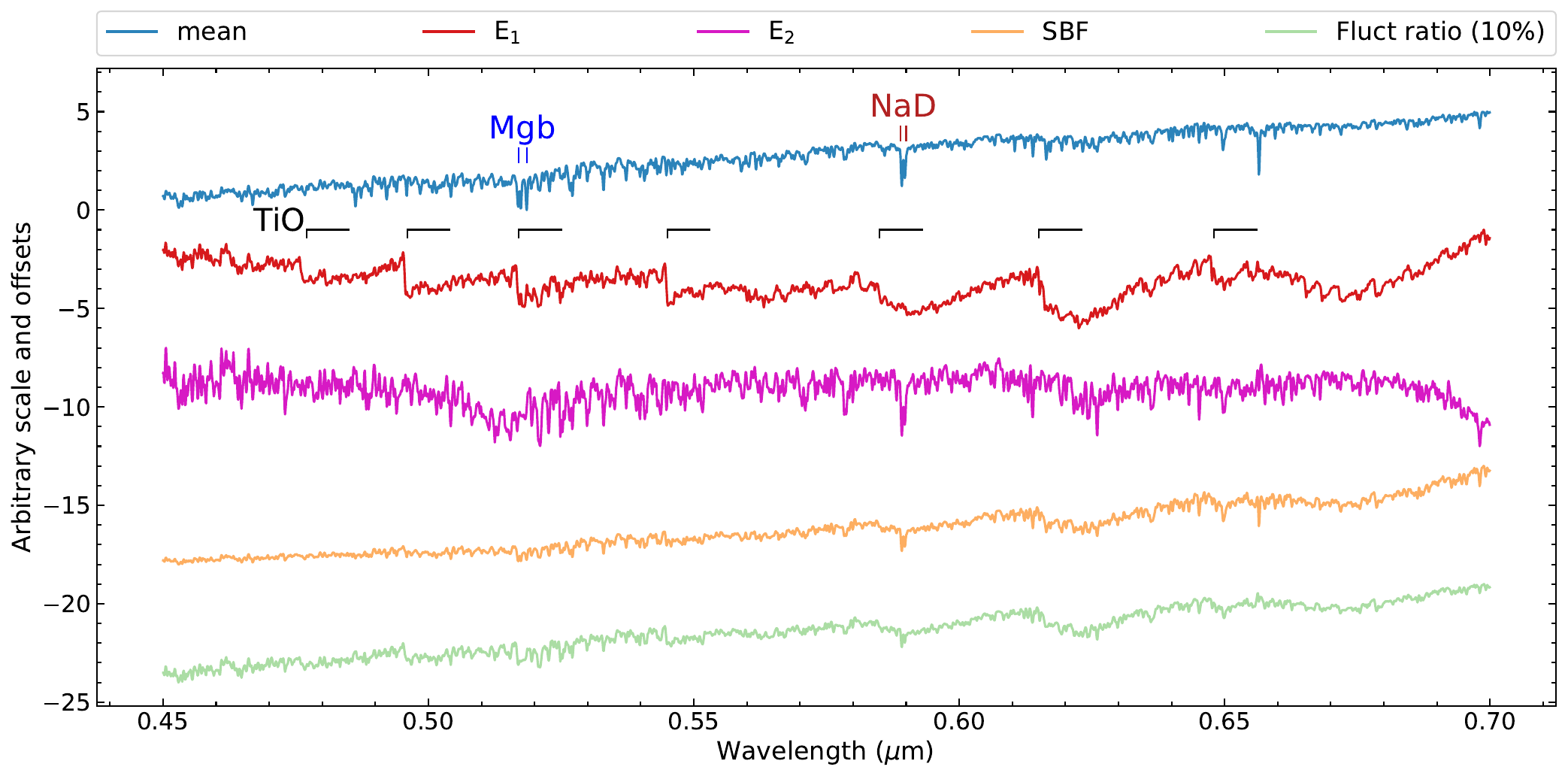}
 \caption{Same methods as in Figure\,\ref{fig:spectral_plot-Iband-t10p00m}. applied to V-band wavelength of solar metallicity and 10 Gyr isochrone.}
 \label{fig:V-spec_iso_ev}
\end{figure*}

Finally, we find somewhat different behaviour in the V band (here actually 0.45--0.70\,$\mu$m, so including parts of B and R). 
In this wavelength range, unlike in the redder bands, the brightest stars are giants of intermediate temperature ($\sim$3700\,K), rather than the very cool stars at the giant-branch tip (see Figure~\ref{fig:main-Vband-t10p00m}). 
As a consequence, the various methods shown in 
Figure\,\ref{fig:V-spec_iso_ev} respond differently and accentuate distinct features unlike in the red or IR. In the first eigenspectrum, the atomic features familiar from the integrated light are suppressed almost completely. 
All of the strong features in $E_1$ are instead Ti\,O bandheads, arising in cool stars with type around M2 type, i.e. $\sim$3500\,K. The impact of such stars on the integrated light spectrum is almost imperceptible.
The second eigenspectrum visually appears to describe mostly narrow features, notably Na\,D; it is tempting to identify the broad trough at 0.51\,$\mu$m as the Mg\,H band, but this band is strong in dwarf stars, so it would not be expected to contribute to the fluctuations. 

The appearance of the SBF spectrum in V  is markedly different to that of $E_1$, to a greater degree than in any other band explored here. As one explanation for this,  recall that the SBF spectrum effectively reflects the brightest star contributing at each wavelength, and is hence expected to trace warmer stars in the optical. Moreover, because it treats each wavelength channel independently, the type of star which is the brightest can drift as a function of wavelength, with ever warmer stars towards shorter wavelengths.  Conversely, the eigenspectra respond to correlated variation between the wavelength channels, rather than treating each channel in isolation. Thus it can happen that if the PCA requires an M-giant-like leading component to account for Ti\,O variation in the red, the correlation structure recovers features from the same spectral types even into the blue. To the extent that the fluctuation ratio spectrum is also capturing correlated information, albeit mediated by the total sample luminosity, it is not surprising that this spectrum appears more similar to $E_1$ than to the SBF spectrum.

To summarise the results of this section, the PCA method has perhaps its greatest utility at shorter wavelengths, where the information it recovers is more distinct from that carried by the integrated light. In all bands, however, the leading component clearly tends to isolate the contributions from the cool-giants and to suppress features from the turn-off and warmer giants, providing a potentially valuable alternative window on unresolved stellar populations.

\section{Age and metallicity effects}
\label{sec:pars}

We now extend our discussion beyond the reference (10 Gyr, solar-metallicity) model to investigate the PCA results for simple stellar populations (SSPs) with varying age (3, 6, 10, and 13 Gyr) and metallicity ([Fe/H] = --0.3, 0.0, and +0.3), based on the same set of isochrones used in the previous section \citep{Bressan2012}. For completeness, we also address the effect of composite populations, illustrated through extremely bimodal test cases.

\begin{figure}
 \includegraphics[width=\columnwidth]{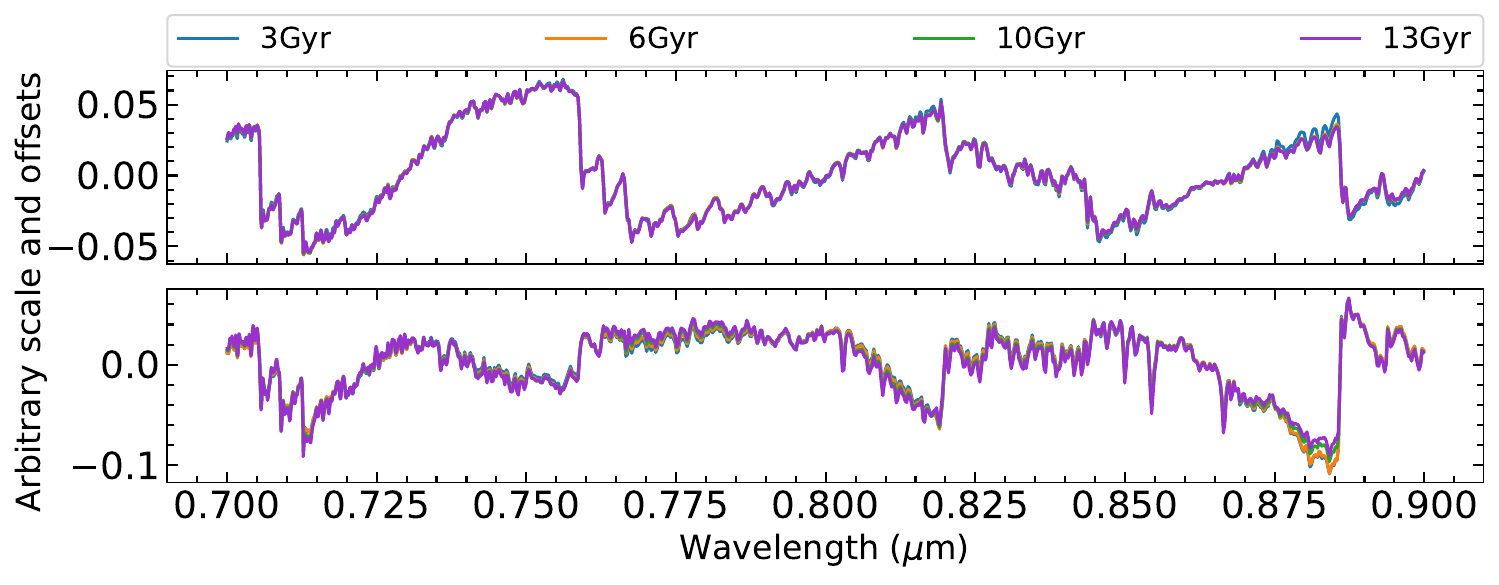}
 \caption{The first (top) and second (bottom) eigenspectra of the spectra in I-band for 10 Gyr solar isochrones with four different ages: 3~Gyr (blue), 6~Gyr (orange), 10~Gyr (green), and 13~Gyr (purple).}
 \label{fig:age-variation_Iband}
\end{figure}

\subsection{Simple populations}

In general, we find that the recovered eigenspectra are insensitive to population age, over the range explored here. Figure~\ref{fig:age-variation_Iband} shows the results for the I-band, but is characteristic of all bands, showing only very minor differences as a function of age. Where small effects are seen, they are not always clearly monotonic with age. This robustness is not unexpected, since the physical parameters of stars near the tip of the RGB and AGB vary little after $\sim$3\,Gyr. The age-variation comparison for the other bands is shown in Appendix\,\ref{apx:ages_comparison_bands}, for completeness. 

\begin{figure}
 \includegraphics[width=\columnwidth]{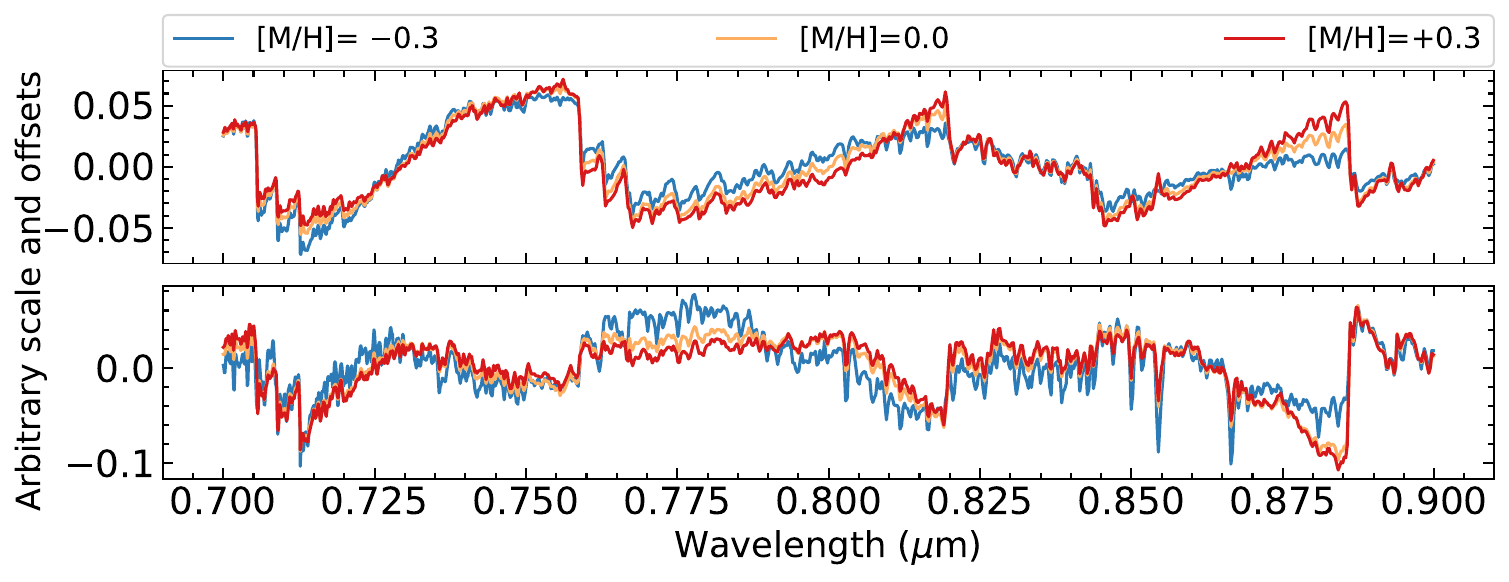}
 \caption{The first (top) and second (bottom) eigenspectra of the spectra in I-band for 10\,Gyr isochrones with three different metallicities: [M/H]\,=\,-0.3 (blue), [M/H]\,=\,+0.0 (orange), and [M/H]\,=\,+0.3 (red).}
 \label{fig:feh-variation_Iband}
\end{figure}

By contrast, metallicity variations can be expected to generate greater differences in the eigenspectra, through changes to the temperature of the giant branches, and hence of the stars driving the fluctuations. 
For the I band, metallicity variations are evident in both egenspectra, $E_1$ and $E_2$, particularly in the continuum, Ti\,O bands, and Ca II triplet region (see Figure\,\ref{fig:feh-variation_Iband}).
The effect on the eigenspectra is monotonic with increasing metallicity. For $E_1$, the dependence can be interpreted as a strengthening of the longer-wavelength bandheads, e.g. $\delta(0-0)$ at 0.885\,$\mu$m, and weakening of the shorter wavelength heads, e.g. $\gamma(\Delta{}v\,=\,0$) at 0.705\,$\mu$m. Qualitatively, this is as expected, since the redder features gain more weight in cooler stars, driving the fluctuations at higher metallicity.

\begin{figure}
 \includegraphics[width=\columnwidth]{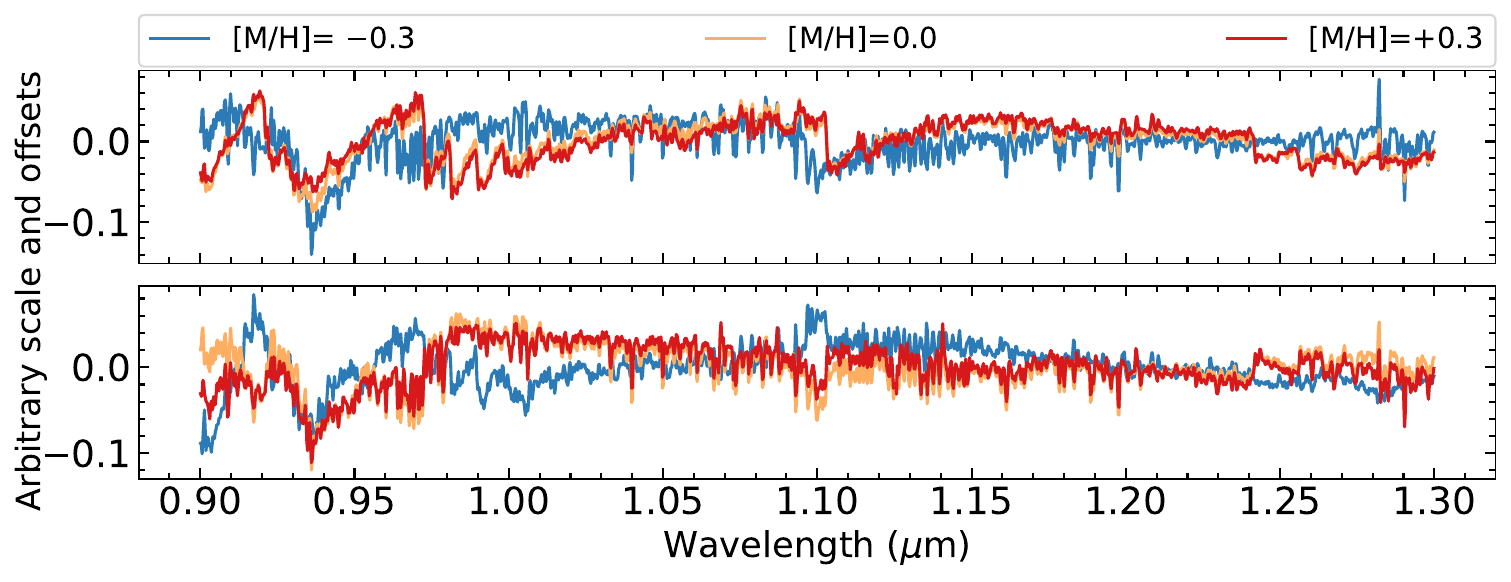}
 \caption{Same as Figure\,\ref{fig:feh-variation_Iband} for the J-band.}
 \label{fig:feh-variation_Jband}
\end{figure}

For the J-band (Figure\,\ref{fig:feh-variation_Jband}), the solar and metal-rich populations yield similar eigenspectra, but notable differences appear in the sub-solar case.
In fact some of the difference is attributable to a ``swapping'' of the first two eigenspectra: features carried by $E_1$ in the [M/H]\,=\,0.0,+0.03 cases (e.g. the 
0.97--1.00\,$\mu$m `sawtooth') are recovered instead in $E_2$ for the metal-poor population, and vice versa. Thus in this case, a different variation component has become the leading term in the PCA recovery.

Finally, the PCA results in the three other bands (V, H, and K) show little sensitivity to metallicity for the 10 Gyr case (see Appendix\,\ref{apx:ages_comparison_bands}), indicating that metallicity variations have a relatively limited impact on the principal components over the abundance range explored.

\subsection{Composite populations}
So far, only SSP models have been considered, but real ETGs are expected at least to have broad metallicity distribution functions, while significant age spreads could also be present in some systems. 
To test for the effects of composite populations, we consider the limiting case of bimodal mixtures, in which stars drawn from two different isochrones are mixed within each realisation. Namely, within each composite realisation, the two populations are normalised to contribute equally to the flux in the band considered, with half of the flux originating from each isochrone. 
Since the luminosity of individual stars depends on the passband, this equal-flux contribution does not correspond to equal numbers of stars or equal stellar masses from the two populations. Consequently, the relative contributions of the two populations in terms of stellar mass and number of stars differ between passbands, such that each passband effectively represents a different composite stellar population.
It may be assumed that if bimodality does not introduce observable effects, then a more continuous distribution in age or metallicity would likewise have limited impact. In all cases, the total luminosity per realisation is kept the same as for the SSP cases previously described.

We again begin by examining the age mixture and then explore the metallicity mixture across all bands. Given the large number of combinations explored -- age mixtures for each metallicity in every band (15 cases in total), and metallicity mixtures for each age in every band (20 cases) -- we restrict the discussion to the overall trends highlighting only a few individual examples. 

\begin{figure}
 \includegraphics[width=\columnwidth]{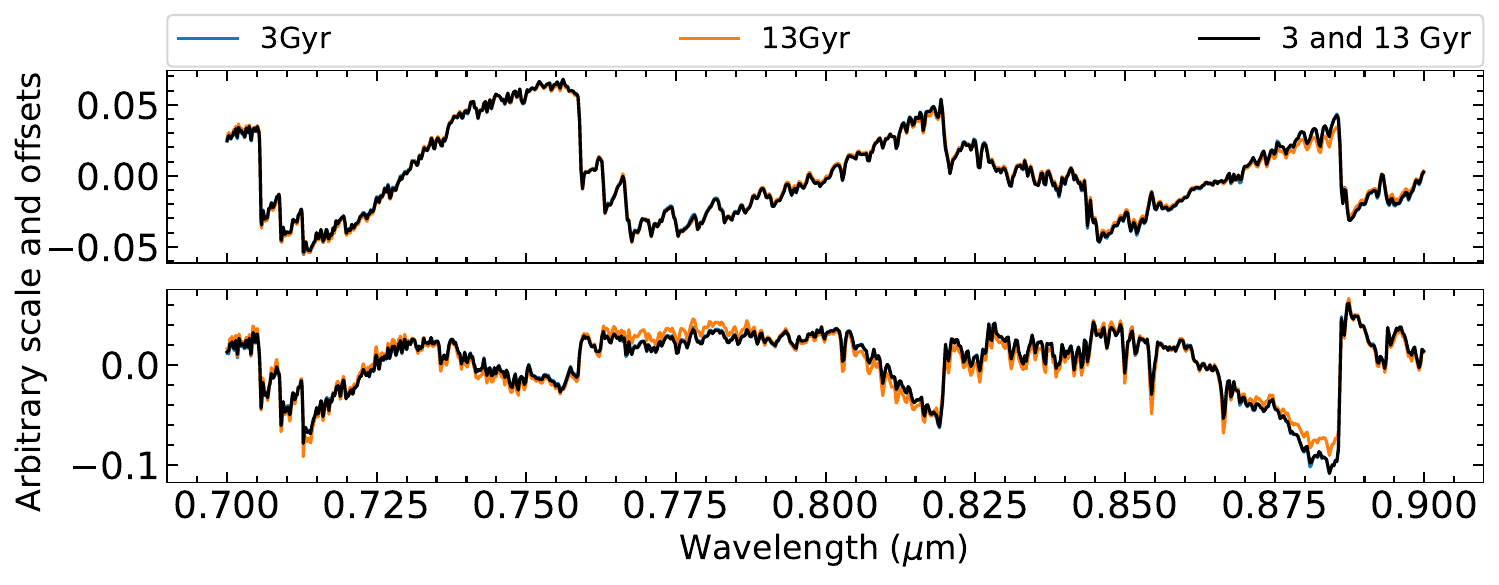}
 \caption{Comparison of SSP-like and composite case for the first (top) and second (bottom) eigenspectra of the spectra in I-band for the solar isochrones. The blue line shows the 3~Gyr SSP case, the orange line the 13~Gyr SSP case, and the black line the mixed 3 and 13~Gyr population case.}
 \label{fig:age-mix-variation_Iband}
\end{figure}

\begin{figure}
 \includegraphics[width=\columnwidth]{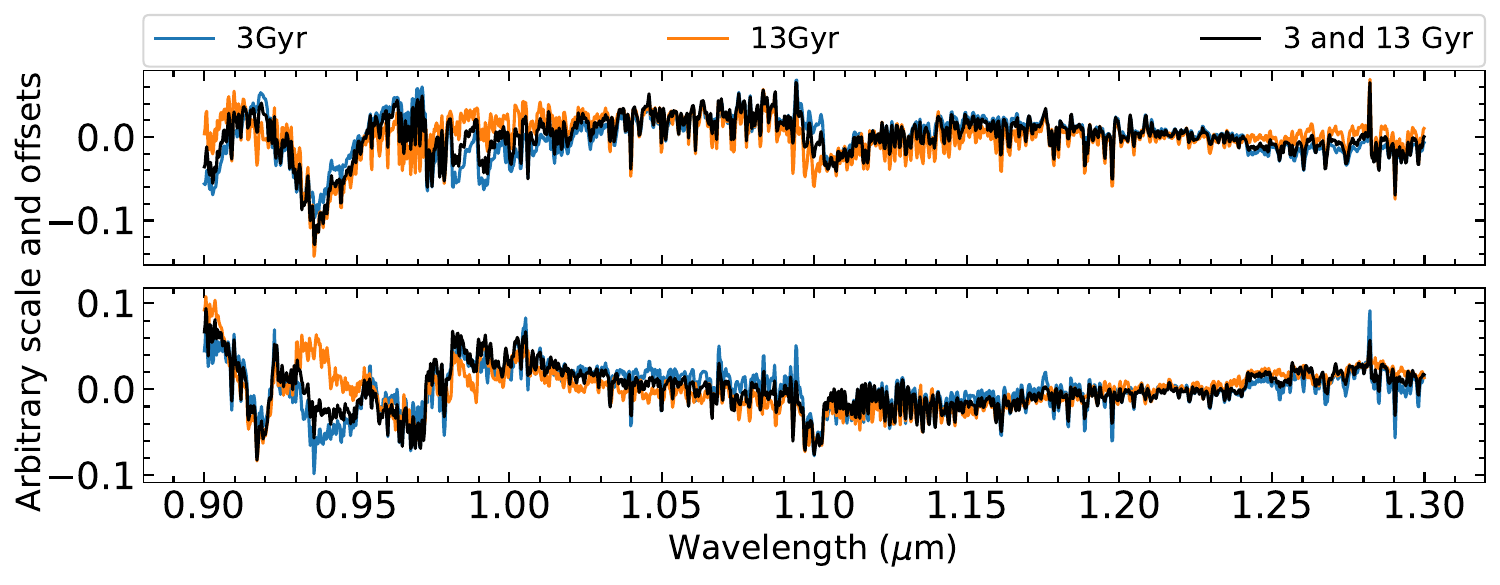}
 \caption{Same as in Figure\,\ref{fig:age-mix-variation_Iband} for the J-band and [M/H]\,=\,--0.3 isochrones.}
 \label{fig:age-mix-variation_Jband_m03}
\end{figure}

\begin{figure}
 \includegraphics[width=\columnwidth]{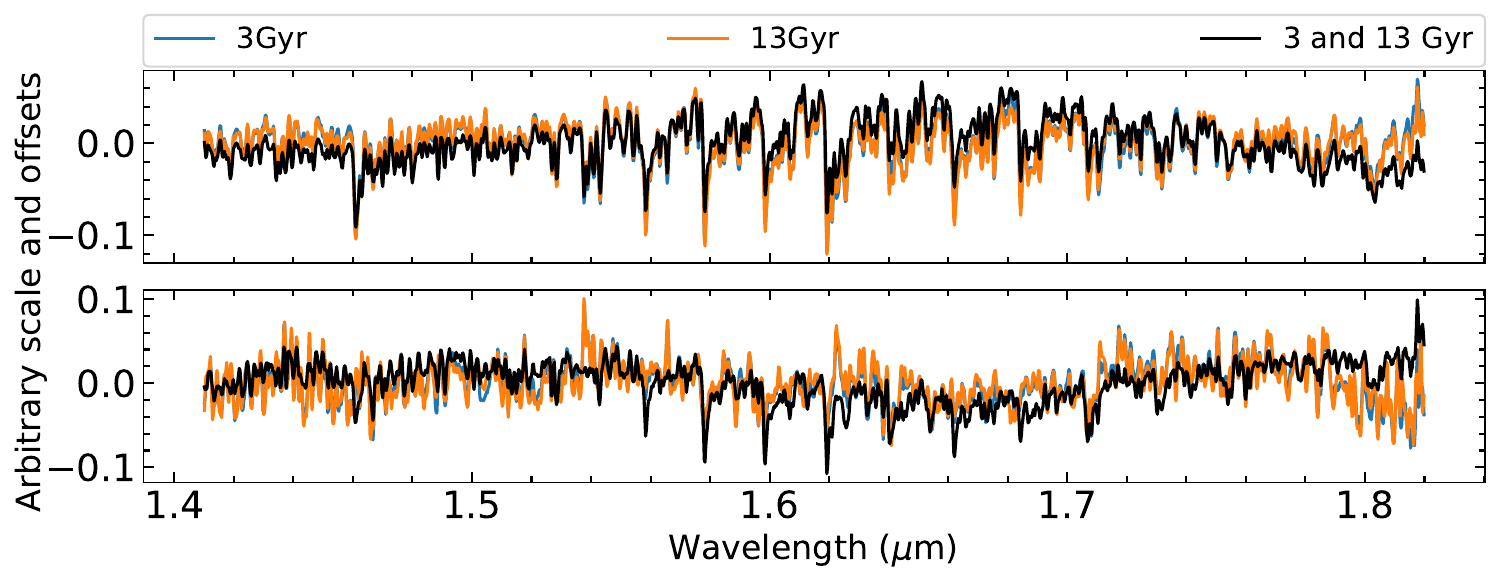}
 \caption{Same as in Figure\,\ref{fig:age-mix-variation_Iband} for the H-band and solar isochrones.}
 \label{fig:age-mix-variation_Hband}
\end{figure}

Similar to the result for SSP cases, where age differences do not manifest in the PCA results, mixing the two extreme ages of 3\,Gyr and 13\,Gyr likewise does not produce strong differences in the eigenspectra. 
In the I band (see Figure\,\ref{fig:age-mix-variation_Iband}), small variations appear for the [M/H]\,=\,0.0 case, with the mixed-age eigenspectrum more closely resembling that of the younger (3\,Gyr) population, though this difference was already modest.
The only case where a sizeable effect is seen is in the metal-poor regime: in the J band, for [M/H]\,=\,--0.3, the 3\,Gyr and 13\,Gyr PCs differ markedly, with the mixed case lying between the two extremes (see Figure\,\ref{fig:age-mix-variation_Jband_m03}). The H band shows the greatest variance across all metallicities, likely due to the shape of the continuum. Figure\,\ref{fig:age-mix-variation_Hband} illustrates this effect, where the mixed case (black line) diverges from the single-SSP examples towards the end of the spectral range. In contrast, the far-blue and far-red bands (V and K) show no clear differences among the 3 Gyr, 13 Gyr, and mixed (3 + 13 Gyr) cases.

\begin{figure}
 \includegraphics[width=\columnwidth]{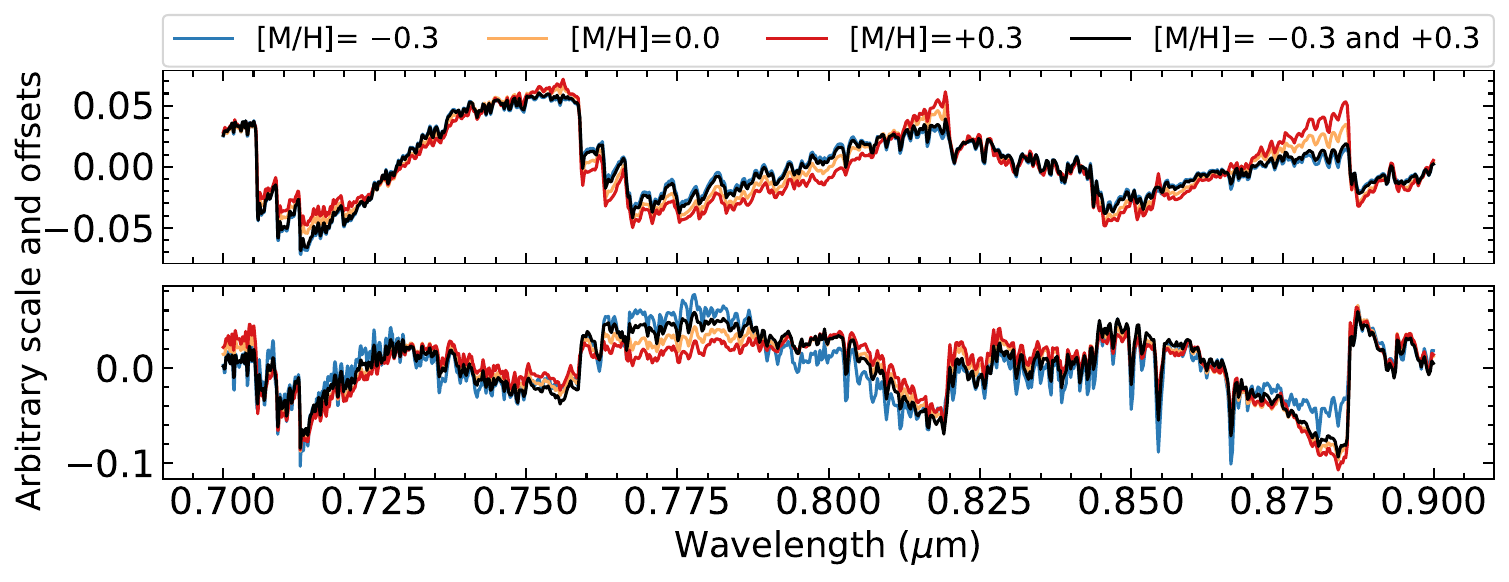}
 \caption{Comparison of SSP-like and composite cases for the first (top) and second (bottom) eigenspectra of the I-band spectra for the 10~Gyr isochrones. The blue line shows the [M/H]\,=\,--0.3 SSP case, the orange line the [M/H]\,=\,0.0 SSP case, the red line the [M/H]\,=\,+0.3 SSP case, and the black line the mixed-metallicity case combining the [M/H]\,=\,--0.3 and +0.3 populations.}
 \label{fig:feh-mix-variation_Iband}
\end{figure}

\begin{figure}
 \includegraphics[width=\columnwidth]{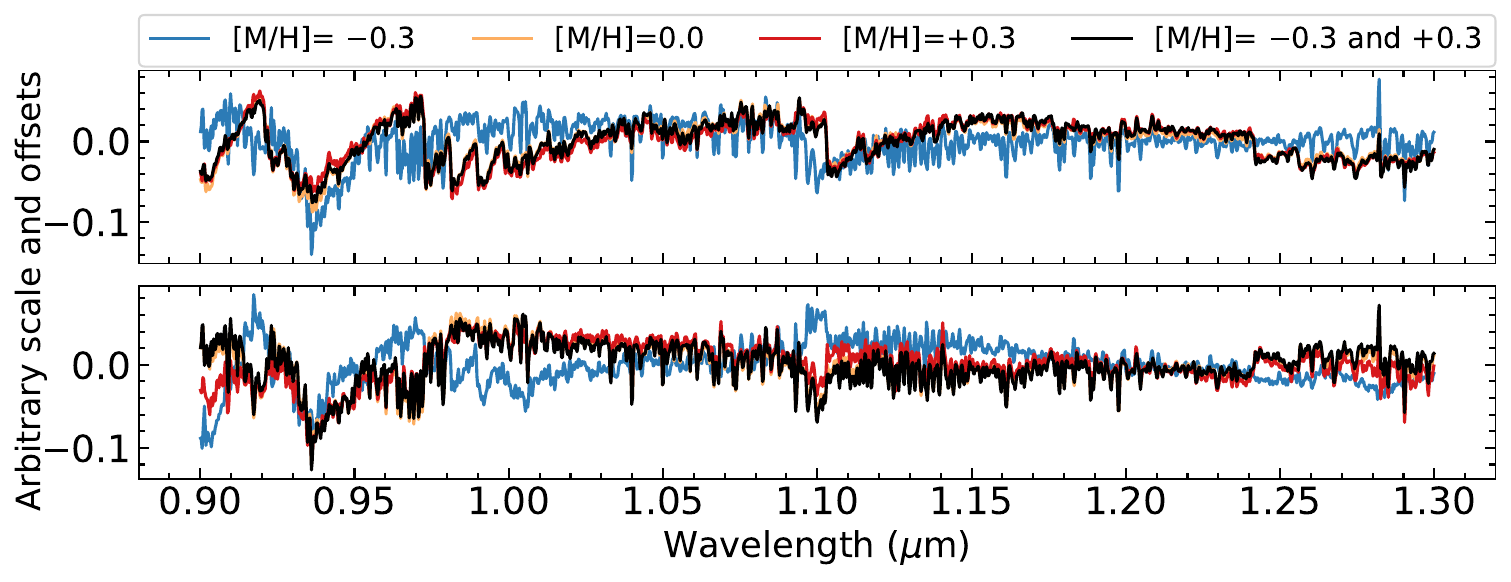}
 \caption{Same as Figure\,\ref{fig:feh-mix-variation_Iband} for the J-band and 10~Gyr isochrones.}
 \label{fig:feh-mix-variation_Jband}
\end{figure}

For the metallicity mixture, we similarly combined the two extreme cases, [M/H]\,=\,--0.3 and [M/H]\,=\,+0.3 dex. Overall, metallicity induces more pronounced variations in the PCA results for the mixed-population case, as it did also in the SSPs.
Figure~\ref{fig:feh-mix-variation_Iband} shows the result in the I band at 10\,Gyr. In $E_1$, we find that the mixed-metallicity population traces the metal-poor case, rather than matching the results for the average, solar, metallicity. 
In the J band, by contrast, (Figure~\ref{fig:feh-mix-variation_Jband}), the mixed-metallicity population more closely follows the [M/H]\,=\,0.0 and [M/H]\,=\,+0.3 SSPs.
The prominent feature at the bluer end of the eigenspectra of the J-band spectra appears to be divided among the first and second PCs differently across the explored metallicities.
This results in noticeable variations in the shapes of the eigenspectra around 0.90-0.95 $\mu$m, with the corresponding percentage contributions reflecting these differences quantitatively (see Appendix\,\ref{App-table}). 
In the H band, the continuum shows a moderate level of variation, with less pronounced differences among individual lines. Overall, across our mixed-population examples, the greatest variations are seen in the J and H bands, whereas the V and I bands are least affected, as this part of the spectrum is not strongly influenced by cool, metal-poor stars.

In summary, the results obtained for composite populations are consistent with those from the SSP analysis. The eigenspectra show minimal variation with age distribution, and slightly more dependence on metallicity distribution, with mixed-metallicity cases generally tending to follow the solar, [M/H]\,=\,0.0, and metal-rich, [M/H]\,=\,+0.3, cases more closely. 
The presence of multiple metallicities does not however, introduce any radical changes to the PCA results. In all cases, the J band, and in particular the 0.9--1.0\,$\mu$m interval, shows more complexity than the other wavelength ranges explored.

\section{Discussion}
\label{sec:disc}

Using Poisson-sampled models, we have shown that the PCA method is able to extract the spectral signatures of cool giant stars from unresolved spectroscopy of stellar populations, over a much wider range in wavelength, and over a broader range in age and metallicity, than previously shown by \stt. 
Across the range of models explored, the results are highly robust in the sense that the eigenspectra capture the spectral signatures of very cool giant stars, regardless of the distribution of ages, and with only modest dependence on metallicity. In general, the stars driving the signal are the most luminous RGB/AGB stars present, though the situation is more complex in the V band. 

The insensitivity of the eigenspectra to age and metallicity means that the PCA method cannot be used to estimate such parameters, which can be better constrained through integrated light spectroscopy. Rather, the potential power of the PCA technique is in testing directly, and in a largely model-independent way, whether the cool giant spectra included in population models are appropriate to the galaxies being observed. As noted in the introduction, this is particularly crucial for massive ETGs, where other evidence indicates differences with respect to the MW, e.g. in abundance ratios and other peculiarities. Although we use models here to guide the interpretation of the eigenspectra, our method can be applied to observations in a purely data-driven manner. As a hypothetical example, if a galaxy were to harbour some new type of giant star with a spectrum different from any in current models, its features would be recovered by the PCA. This power is demonstrated by \cite{2026arXiv260521588S}, where PCA applied to real observations isolates carbon-star signatures in NGC 5102 and reveals evidence for an additional cool giant component, possibly associated with long-period variables, in NGC 5128.
The model predictions explored in the present paper establish a baseline against which such results can be evaluated.

Compared to the SBF spectrum discussed by  \cite{Mitzkus2018}, the PCA approach shows greater sensitivity to the coolest stars across the full wavelength explored. The difference between the two techniques is particularly noticeable in the optical. Because SBF traces the most luminous contributors to the light at each wavelength, the dominant contributors vary with wavelength, trending towards warmer stars in the blue. PCA instead isolates coherent spectral signatures associated with specific stellar types, even when they contribute very little to the integrated light. In the optical, especially, this allows our technique to access qualitatively distinct information from observational data. 
Provided in Table\,\ref{tab:basecases-stat} $\bar{N}$ and N$_\text{ev}$ roughly agree in the blue end of the spectrum, however, towards the K-band $\bar{N}$ appears to underestimate the contribution of the evolved stars.
The reason being, that in the K band the fluctuations are dominated by stars at the very tip of the isochrone, so a small number of exceptionally luminous stars can dominate the signal, resulting in a relatively low $\bar{N}$. In the V band, the most luminous contributors are not the coolest stars at the tip of the isochrone (see Figure\,\ref{fig:main-Vband-t10p00m}), and the fluctuations are driven by a different part of the stellar population.

Among the promising scientific applications of our approach would be to test possible explanations for the long-standing discrepancies between stellar population models and observations of the CO band strengths in ETGs \citep{LaBarbera2025}.
By isolating the spectra of the coolest giant contributions directly from unresolved data, PCA provides a means to test whether such features originate predominantly from giant or dwarf stars, thereby helping to disentangle competing interpretations and to refine population synthesis models. 
For example, if the enhanced CO originates preferentially in the most luminous giants, it will be enhanced in the fluctuation eigenspectrum, compared to the integrated light. By contrast, if this signal affects only lower-luminosity stars, it will be greatly suppressed.

The PCA method might also help assess the reality of the bottom-heavy IMF signature in ETGs. 
The eigenspectra are not directly sensitive to the IMF, 
because the Poisson fluctuations are driven by stars in a very narrow mass range, for SSP-like populations.
However, as noted in the introduction,  integrated-light modelling of ETGs (e.g. \citealt{Conroy2012, LaBarbera2025}, etc.) is quite sensitive to the treatment of cool giants, which collectively contribute far more light than the low-mass dwarfs that carry the IMF signal. Therefore, any mismatch between models and observations revealed through PCA would directly inform the IMF question via its impact on the representation of cool giant stars.

In this study, we focused on populations older than $\sim$3~Gyr, which are typically dominated by a single bright spectral component shaped by oxygen-rich M-type giants, yielding a primary PCA component associated with these luminous stars.
Extending the method to younger populations ($\lesssim$3~Gyr), the method is able to recover separately the distinct contributions from both oxygen-rich and carbon-rich giants on the thermally-pulsing AGB (TP-AGB), as we have demonstrated empirically with H-band observations of a post-starburst galaxy \citep{2026arXiv260521588S}.
Modelling such younger populations remains challenging because the tracks and spectra for TP-AGB stars, and the transition to carbon-rich atmospheres, are complex and differ widely among modelling groups. 
Empirical PCA-based approaches that disentangle these stellar components should thus provide a valuable avenue for characterising the stellar content and evolutionary histories of ETGs at different ages.

From an observational perspective, the practical requirement for applying this method is a sufficiently large number of independent spatial samples drawn from a region that can be considered uniform in its underlying stellar content rather than high surface brightness or near-resolved stellar populations.
While the signal-to-noise of the fluctuations is independent of surface brightness and pixel size in an idealised sense, real observations are limited by the need to avoid systematic variations across the field of view and by the requirement that Poisson fluctuations from individual stars exceed other noise sources. In practice, this drives observations toward high spatial resolution, which allows many independent samples to be obtained within a small, homogeneous region, and toward regimes where the number of photons per star is sufficient to detect the fluctuations.

The practical implementation of this technique is tightly linked to the capabilities of current and upcoming facilities. High–spatial-resolution IFU instruments on the ELT, in particular HARMONI \citep{Thatte2022}, are ideally suited to applying this technique to the dense central regions of massive and distant early-type galaxies. At optical wavelengths, MAVIS (VLT) will provide IFU capabilities over a much wider sky area than current adaptive-optics systems  \citep{2020arXiv200909242M}, enabling similar analyses in nearby systems without the stringent guide-star requirements of existing facilities. In parallel, \textit{JWST} offers a complementary, space-based route to probing stellar population fluctuations at low background and high sensitivity in the near-infrared.

\section{Conclusions}
\label{sec:concs}
We have extended and demonstrated the advantages of the PCA method for analysing integral-field spectroscopic data of ETGs. We have also compared this approach with other commonly used methods, such as SBF and fluctuation spectra \citep{Mitzkus2018, vanDokkum2014}. In this work, we have demonstrated the effectiveness and robustness of the PCA method across a wide range of physical parameters, including age, metallicity, and wavelength.

Our main conclusions are as follows:
\begin{itemize}
\item[--] In general the spectral variation among samples is dominated  ($\gtrsim$80\% of total variance) by a single PC, representing the contribution of cool giant stars, which are usually the most luminous in the population. A second component (typically 5--20\%, but larger in J-band) captures variation `along' the giant branch. 
 \item[--] The eigenspectra are particularly effective in isolating cool giants rather than simply tracing the brightest stars in the population.    In particular, the PCA approach demonstrates enhanced sensitivity to cool giants in the blue and optical ranges compared to SBF spectra.
    \item[--] The fluctuations identified by PCA primarily reflect intrinsic spectral variations within a narrow mass range (RGB and early AGB). The properties of these stars vary somewhat with metallicity, but little with population age, so that the PCA is insensitive to age, or the IMF.

    \item[--] This method can be directly applied to observed IFS data of ETGs to empirically test stellar population models, particularly in regions where discrepancies, such as the CO band strength, remain unexplained. Extending the technique to truly massive elliptical galaxies (all at distances $\gtrsim$\,10\,Mpc) will require high-precision data with high spatial resolution, and is therefore a natural application for future facilities such as ELT/HARMONI.
    
\end{itemize}

More broadly, compared to other techniques commonly applied to unresolved spectroscopy, PCA provides a direct and flexible means of isolating and characterising the contribution of cool giants to integrated spectra. By decomposing pixel-to-pixel spectral variations into a small number of orthogonal components, the PCA framework offers a data-driven route to disentangle stellar contributions in unresolved systems. This approach complements model-based methods and opens new avenues for studying stellar population gradients and galaxy evolution. Future work will extend this framework towards younger stellar populations, where the presence of luminous asymptotic giant branch stars and carbon-rich giants is expected to produce additional, distinct fluctuation signatures. Testing these predictions observationally with new high-resolution integral-field spectroscopy will be an important step towards establishing the full diagnostic power of the method.

\section*{Acknowledgements}

Both authors were supported by the STFC through consolidated grant ST/X001075/1.
We thank the anonymous referee for a careful and critical reading of the manuscript.

\section*{Data Availability}

The data products underlying this work, including the PCA-ready stellar population spectra and associated files required to reproduce the analysis, are available at the GitHub repository: https://github.com/agvozdenko/stellar-population-pca. These data may be used to reproduce the PCA analysis presented in this work, or to implement alternative treatments. Further similar datasets can be made available upon request to the first author.

The stellar population models are based on Poisson-sampled realisations of the underlying isochrone distribution. Consequently, the data products released through the repository are not identical to the specific stochastic realisations used to generate the numerical results presented in this paper. Derived quantities obtained from the released files may therefore differ from the values reported herein, reflecting the expected stochastic variations introduced by Poisson sampling. The qualitative conclusions and overall behaviour of the models are unaffected by these realisation-to-realisation differences.



\bibliographystyle{mnras}
\bibliography{bibl}

@ARTICLE{Smith2022,
       author = {{Smith}, Russell J.},
        title = "{Probing cool giants in unresolved galaxies using fluctuation eigenspectra: A demonstration using high-resolution MUSE observations of NGC 5128}",
      journal = {\mnras},
         year = 2022,
        month = feb,
       volume = {509},
       number = {4},
        pages = {5737-5746},
          doi = {10.1093/mnras/stab3415},
archivePrefix = {arXiv},
       eprint = {2111.09317},
 primaryClass = {astro-ph.GA},
       adsurl = {https://ui.adsabs.harvard.edu/abs/2022MNRAS.509.5737S}
}

@ARTICLE{Conroy2012,
       author = {{Conroy}, Charlie and {van Dokkum}, Pieter G.},
        title = "{The Stellar Initial Mass Function in Early-type Galaxies From Absorption Line Spectroscopy. II. Results}",
      journal = {\apj},
         year = 2012,
        month = nov,
       volume = {760},
       number = {1},
          eid = {71},
        pages = {71},
          doi = {10.1088/0004-637X/760/1/71},
archivePrefix = {arXiv},
       eprint = {1205.6473},
 primaryClass = {astro-ph.CO},
       adsurl = {https://ui.adsabs.harvard.edu/abs/2012ApJ...760...71C}
}

@ARTICLE{2026arXiv260521588S,
       author = {{Smith}, Russell J. and {Gvozdenko}, Anastasia},
        title = "{Decomposing cool stellar populations with H-band spectral fluctuations: Long-period variable stars in NGC 5128 and carbon stars in NGC 5102}",
      journal = {arXiv e-prints},
         year = 2026,
        month = may,
          eid = {arXiv:2605.21588},
        pages = {arXiv:2605.21588},
          doi = {10.48550/arXiv.2605.21588},
archivePrefix = {arXiv},
       eprint = {2605.21588},
 primaryClass = {astro-ph.GA},
       adsurl = {https://ui.adsabs.harvard.edu/abs/2026arXiv260521588S}
}

@ARTICLE{1995AJ....110.1071C,
       author = {{Connolly}, A.~J. and {Szalay}, A.~S. and {Bershady}, M.~A. and {Kinney}, A.~L. and {Calzetti}, D.},
        title = "{Spectral Classification of Galaxies: an Orthogonal Approach}",
      journal = {\aj},
         year = 1995,
        month = sep,
       volume = {110},
        pages = {1071},
          doi = {10.1086/117587},
archivePrefix = {arXiv},
       eprint = {astro-ph/9411044},
 primaryClass = {astro-ph},
       adsurl = {https://ui.adsabs.harvard.edu/abs/1995AJ....110.1071C}
}

@ARTICLE{1992ApJ...398..476F,
       author = {{Francis}, Paul J. and {Hewett}, Paul C. and {Foltz}, Craig B. and {Chaffee}, Frederic H.},
        title = "{An Objective Classification Scheme for QSO Spectra}",
      journal = {\apj},
         year = 1992,
        month = oct,
       volume = {398},
        pages = {476},
          doi = {10.1086/171870},
       adsurl = {https://ui.adsabs.harvard.edu/abs/1992ApJ...398..476F}
}

@ARTICLE{2007MNRAS.381..543W,
       author = {{Wild}, Vivienne and {Kauffmann}, Guinevere and {Heckman}, Tim and {Charlot}, St{\'e}phane and {Lemson}, Gerard and {Brinchmann}, Jarle and {Reichard}, Tim and {Pasquali}, Anna},
        title = "{Bursty stellar populations and obscured active galactic nuclei in galaxy bulges}",
      journal = {\mnras},
         year = 2007,
        month = oct,
       volume = {381},
       number = {2},
        pages = {543-572},
          doi = {10.1111/j.1365-2966.2007.12256.x},
archivePrefix = {arXiv},
       eprint = {0706.3113},
 primaryClass = {astro-ph},
       adsurl = {https://ui.adsabs.harvard.edu/abs/2007MNRAS.381..543W}
}

@ARTICLE{1998MNRAS.298..361B,
       author = {{Bailer-Jones}, Coryn A.~L. and {Irwin}, Mike and {von Hippel}, Ted},
        title = "{Automated classification of stellar spectra - II. Two-dimensional classification with neural networks and principal components analysis}",
      journal = {\mnras},
         year = 1998,
        month = aug,
       volume = {298},
       number = {2},
        pages = {361-377},
          doi = {10.1046/j.1365-8711.1998.01596.x},
archivePrefix = {arXiv},
       eprint = {astro-ph/9803050},
 primaryClass = {astro-ph},
       adsurl = {https://ui.adsabs.harvard.edu/abs/1998MNRAS.298..361B}
}

@ARTICLE{Lancon2000,
       author = {{Lan{\c{c}}on}, A. and {Wood}, P.~R.},
        title = "{A library of 0.5 to 2.5 mu m spectra of luminous cool stars}",
      journal = {\aaps},
         year = 2000,
        month = oct,
       volume = {146},
        pages = {217-249},
          doi = {10.1051/aas:2000269},
       adsurl = {https://ui.adsabs.harvard.edu/abs/2000A&AS..146..217L}
}

@ARTICLE{Joyce1998,
       author = {{Joyce}, Richard R. and {Hinkle}, Kenneth H. and {Wallace}, Lloyd and {Dulick}, Michael and {Lambert}, David L.},
        title = "{Spectra of Cool Stars in the J Band (1.0-1.3 {\ensuremath{\mu}}m) at Medium Resolution}",
      journal = {\aj},
         year = 1998,
        month = nov,
       volume = {116},
       number = {5},
        pages = {2520-2529},
          doi = {10.1086/300594},
       adsurl = {https://ui.adsabs.harvard.edu/abs/1998AJ....116.2520J}
}

@INPROCEEDINGS{Thatte2022,
       author = {{Thatte}, Niranjan A. and {Melotte}, Dave and {Neichel}, Benoit and {Le Mignant}, David and {Bryson}, Ian and {Clarke}, Fraser and {Ferraro-Wood}, Vanessa and {Fusco}, Thierry and {Gonzalez}, Oscar and {Schnetler}, Hermine and {Tecza}, Matthias and {Wilson}, Sandi and {{\'A}lvarez Urue{\~n}a}, Alonso and {Vilaseca}, Heribert A. and {Arribas}, Santiago and {Carracedo Carballale}, Gonzalo Jos{\'e} and {Crespo}, Alejandro and {Estrada Piqueras}, Alberto and {Garc{\'\i}a Garc{\'\i}a}, Miriam and {Mart{\'\i}nez Mart{\'\i}n}, Cecilai and {Pereira Santaella}, Miguel and {Perna}, Michele and {Piqueras Lopez}, Javier and {Bouch{\'e}}, Niolas and {Boudon}, Didier and {Daguise}, Eric and {Disseau}, Karen and {Fensch}, J{\'e}r{\'e}my J. and {Girardot}, Adrien and {Guibert}, Matthieu and {Jarno}, Aur{\'e}lien and {Jeanneau}, Alexandre and {Krogager}, Jens-Kristian and {Laurent}, Florence and {Loupias}, Magali and {Migniau}, Jean-Emmanuel and {Piqueras}, Laure and {Remillieux}, Alban and {Richard}, Johan and {Pecontal}, Arlette and {Bardou}, Lisa F. and {Barr}, David and {Cetre}, Sylvain and {Deshmukh}, Rishi and {Dimoudi}, Sofia and {Dubbledam}, Marc and {Dunn}, Andrew and {Gadotti}, Dimitri and {Guy}, Joss J. and {King}, David and {Little}, David J. and {McLeod}, Anna and {Morris}, Simon and {Morris}, Tim and {O'Brien}, Kieran S. and {Ronson}, Emily and {Smith}, Russell and {Staykov}, Lazar and {Swinbank}, Mark and {Townson}, Matthew and {Accardo}, Matteo and {Alvarez Mendez}, Domingo and {George}, Elizabeth and {Hopgood}, Joshua and {Ives}, Derek and {Mehrgan}, Leander and {Mueller}, Eric and {Reyes}, Javier and {Conzelmann}, Ralf and {Gutierrez Cheetham}, Pablo and {Alonso-S{\'a}nchez}, {\'A}ngel and {Battaglia}, Giuseppina and {Cagigas}, Miguel and {Chulani}, Haresh and {Delgado Garc{\'\i}a}, Graciela and {Fernandez Izquierdo}, Patricia and {Fragoso L{\'o}pez}, Ana Bel{\'e}n. and {Garc{\'\i}a-Lorenzo}, Bego{\~n}a. and {Hern{\'a}ndez Gonz{\'a}lez}, Alberto and {Hernandez Suarez}, Elvio and {Herreros Linares}, Jose Miguel and {Joven}, Enrique and {L{\'o}pez L{\'o}pez}, Roberto and {Lujan Gonzalez}, Alejandro Antonio and {Mart{\'\i}n Hernando}, Yolanda and {Mediavilla}, Evencio and {Men{\'e}ndez Mendoza}, Sa{\'u}l and {Montoya Mart{\'\i}nez}, Luz Maria and {Pe{\~n}ate Castro}, Jos{\'e} and {P{\'e}rez}, {\'A}lvaro and {Rasilla}, Jose Luis and {Rebolo}, Rafael and {Rodr{\'\i}guez Ramos}, Luis Fernando and {Vega-Moreno}, Afrodisio and {Viera}, Teodora and {Zanon Dametto}, Natacha and {Carlotti}, Alexis and {Correia}, Jean-Jacques and {Curaba}, St{\'e}phane and {Delboulbe}, Alain and {Guieu}, Sylvain and {Hours}, Adrien and {Hubert}, Zoltan and {Jocou}, Laurent and {Magnard}, Yves and {Moulin}, Thibaut and {Pancher}, Fabrice and {Rabou}, Patrick and {Stadler}, Eric and {V{\'e}rove}, Maxime and {Contini}, Thierry and {Larrieu}, Marie and {Boebion}, Olivier and {Fantei-Caujolle}, Yan and {Lecron}, Daniel and {Rousseau}, Sylvain and {Amram}, Philippe and {Beltramo-Martin}, Olivier and {Bon}, William and {Bonnefoi}, Anne and {Ceria}, William and {Challita}, Zalpha and {Charles}, Yannick and {Choquet}, Elodie and {Correia}, Carlos and {Costille}, Anne and {Dohlen}, Kjetil and {Ducret}, Franck and {El Hadi}, Kacem and {Gach}, Jean-Luc and {Gimenez}, Jean-Luc and {Groussin}, Olivier and {Jaquet}, Marc and {Jouve}, Pierre and {Madec}, Fabrice and {Pedreros}, Felipe and {Renault}, Edgard and {Sanchez}, Patrice and {Vigan}, Arthur and {Vola}, Pascal and {Zavago}, Annie and {Fetick}, Romain and {Lim}, Caroline and {Petit}, Cyril and {Sauvage}, Jean-Francois and {V{\'e}drenne}, Nicolas and {Bagci}, Fehim Taha and {Caldwell}, Martin E. and {Elliott}, Ellis and {Hiscock}, Peter and {Johnson}, Emma and {Nalagatla}, Murali and {Seitis}, Aristea and {Wells}, Mark and {Black}, Martin and {Bond}, Charlotte and {Brierley}, Saskia and {Campbell}, Kenny and {Campbell}, Neil and {Carruthers}, James and {Cochrane}, William and {Evans}, Chris and {Harman}, Joel and {Humphreys}, William and {Louth}, Thomas and {Miller}, Chris and {Montgomery}, David and {Murali}, Meenu and {Murray}, John and {O'Malley}, Norman and {Sanchez-Janssen}, Ruben and {Schwartz}, Noah and {Smith}, Patrick and {Strachan}, Jonathan and {Todd}, Stephen and {Watt}, Stuart and {Wells}, Martyn and {Yaqoob}, Asim and {Bell}, Eric and {Gnedin}, Oleg O. and {Gultekin}, Kayhan and {Mateo}, Mario and {Meyer}, Michael and {Ahmad}, Munadi and {Birkby}, Jayne and {Booth}, Michael and {Cappellari}, Michele and {Castillo Dominguez}, Edgar and {Chao Ortiz}, Jorge and {Gooding}, David and {Grisdale}, Kearn and {Hidalgo}, Andrea and {Hogan}, Laurence and {Kariuki}, James and {Lewis}, Ian and {Lowe}, Adam and {Ozer}, Zeynep and {Routledge}, Laurence and {Rigopoulou}, Dimitra and {York}, Alec},
        title = "{HARMONI at ELT: overview of the capabilities and expected performance of the ELT's first light, adaptive optics assisted integral field spectrograph.}",
    booktitle = {Ground-based and Airborne Instrumentation for Astronomy IX},
         year = 2022,
       editor = {{Evans}, Christopher J. and {Bryant}, Julia J. and {Motohara}, Kentaro},
       series = {Society of Photo-Optical Instrumentation Engineers (SPIE) Conference Series},
       volume = {12184},
        month = aug,
          eid = {1218420},
        pages = {1218420},
          doi = {10.1117/12.2628834},
       adsurl = {https://ui.adsabs.harvard.edu/abs/2022SPIE12184E..20T}
}

@ARTICLE{Mitzkus2018,
       author = {{Mitzkus}, Martin and {Walcher}, C. Jakob and {Roth}, Martin M. and {Coelho}, Paula R.~T. and {Cioni}, Maria-Rosa L. and {Raimondo}, Gabriella and {Rejkuba}, Marina},
        title = "{Surface brightness fluctuation spectrum: a new probe of evolved stars in unresolved stellar populations}",
      journal = {\mnras},
         year = 2018,
        month = oct,
       volume = {480},
       number = {1},
        pages = {629-651},
          doi = {10.1093/mnras/sty1840},
       adsurl = {https://ui.adsabs.harvard.edu/abs/2018MNRAS.480..629M}
}

@ARTICLE{2012ApJ...747...69C,
       author = {{Conroy}, Charlie and {van Dokkum}, Pieter},
        title = "{Counting Low-mass Stars in Integrated Light}",
      journal = {\apj},
         year = 2012,
        month = mar,
       volume = {747},
       number = {1},
          eid = {69},
        pages = {69},
          doi = {10.1088/0004-637X/747/1/69},
archivePrefix = {arXiv},
       eprint = {1109.0007},
 primaryClass = {astro-ph.CO},
       adsurl = {https://ui.adsabs.harvard.edu/abs/2012ApJ...747...69C}
}

@ARTICLE{2017MNRAS.464.3597L,
       author = {{La Barbera}, F. and {Vazdekis}, A. and {Ferreras}, I. and {Pasquali}, A. and {Allende Prieto}, C. and {R{\"o}ck}, B. and {Aguado}, D.~S. and {Peletier}, R.~F.},
        title = "{IMF and [Na/Fe] abundance ratios from optical and NIR spectral features in early-type galaxies}",
      journal = {\mnras},
         year = 2017,
        month = jan,
       volume = {464},
       number = {3},
        pages = {3597-3616},
          doi = {10.1093/mnras/stw2407},
archivePrefix = {arXiv},
       eprint = {1610.03853},
 primaryClass = {astro-ph.GA},
       adsurl = {https://ui.adsabs.harvard.edu/abs/2017MNRAS.464.3597L}
}

@ARTICLE{2015A&A...578A.117C,
       author = {{Chantereau}, W. and {Charbonnel}, C. and {Decressin}, T.},
        title = "{Evolution of long-lived globular cluster stars. I. Grid of stellar models with helium enhancement at [Fe/H] = -1.75}",
      journal = {\aap},
         year = 2015,
        month = jun,
       volume = {578},
          eid = {A117},
        pages = {A117},
          doi = {10.1051/0004-6361/201525929},
archivePrefix = {arXiv},
       eprint = {1504.01878},
 primaryClass = {astro-ph.SR},
       adsurl = {https://ui.adsabs.harvard.edu/abs/2015A&A...578A.117C}
}

@ARTICLE{1990ApJ...364...35G,
       author = {{Greggio}, Laura and {Renzini}, Alvio},
        title = "{Clues on the Hot Star Content and the Ultraviolet Output of Elliptical Galaxies}",
      journal = {\apj},
         year = 1990,
        month = nov,
       volume = {364},
        pages = {35},
          doi = {10.1086/169384},
       adsurl = {https://ui.adsabs.harvard.edu/abs/1990ApJ...364...35G}
}

@ARTICLE{1999ARA&A..37..603O,
       author = {{O'Connell}, Robert W.},
        title = "{Far-Ultraviolet Radiation from Elliptical Galaxies}",
      journal = {\araa},
         year = 1999,
        month = jan,
       volume = {37},
        pages = {603-648},
          doi = {10.1146/annurev.astro.37.1.603},
archivePrefix = {arXiv},
       eprint = {astro-ph/9906068},
 primaryClass = {astro-ph},
       adsurl = {https://ui.adsabs.harvard.edu/abs/1999ARA&A..37..603O}
}

@ARTICLE{2023MNRAS.523.3450K,
       author = {{Knowles}, Adam T. and {Sansom}, A.~E. and {Vazdekis}, A. and {Allende Prieto}, C.},
        title = "{sMILES SSPs: a library of semi-empirical MILES stellar population models with variable [{\ensuremath{\alpha}}/Fe] abundances}",
      journal = {\mnras},
         year = 2023,
        month = aug,
       volume = {523},
       number = {3},
        pages = {3450-3470},
          doi = {10.1093/mnras/stad1647},
archivePrefix = {arXiv},
       eprint = {2306.05942},
 primaryClass = {astro-ph.GA},
       adsurl = {https://ui.adsabs.harvard.edu/abs/2023MNRAS.523.3450K}
}

@ARTICLE{2005ApJ...621..673T,
       author = {{Thomas}, Daniel and {Maraston}, Claudia and {Bender}, Ralf and {Mendes de Oliveira}, Claudia},
        title = "{The Epochs of Early-Type Galaxy Formation as a Function of Environment}",
      journal = {\apj},
         year = 2005,
        month = mar,
       volume = {621},
       number = {2},
        pages = {673-694},
          doi = {10.1086/426932},
archivePrefix = {arXiv},
       eprint = {astro-ph/0410209},
 primaryClass = {astro-ph},
       adsurl = {https://ui.adsabs.harvard.edu/abs/2005ApJ...621..673T}
}

@ARTICLE{2013MNRAS.429L..15F,
       author = {{Ferreras}, I. and {La Barbera}, F. and {de La Rosa}, I.~G. and {Vazdekis}, A. and {de Carvalho}, R.~R. and {Falcon-Barroso}, J. and {Ricciardelli}, E.},
        title = "{Systematic variation of the stellar initial mass function with velocity  dispersion in early-type galaxies.}",
      journal = {\mnras},
         year = 2013,
        month = feb,
       volume = {429},
        pages = {L15-L19},
          doi = {10.1093/mnrasl/sls014},
archivePrefix = {arXiv},
       eprint = {1206.1594},
 primaryClass = {astro-ph.CO},
       adsurl = {https://ui.adsabs.harvard.edu/abs/2013MNRAS.429L..15F}
}

@ARTICLE{2014ApJ...780...33C,
       author = {{Conroy}, Charlie and {Graves}, Genevieve J. and {van Dokkum}, Pieter G.},
        title = "{Early-type Galaxy Archeology: Ages, Abundance Ratios, and Effective Temperatures from Full-spectrum Fitting}",
      journal = {\apj},
         year = 2014,
        month = jan,
       volume = {780},
       number = {1},
          eid = {33},
        pages = {33},
          doi = {10.1088/0004-637X/780/1/33},
archivePrefix = {arXiv},
       eprint = {1303.6629},
 primaryClass = {astro-ph.CO},
       adsurl = {https://ui.adsabs.harvard.edu/abs/2014ApJ...780...33C}
}

@ARTICLE{1998PASP..110..888W,
       author = {{Worthey}, Guy},
        title = "{Abundance Ratio Trends and Nucleosynthesis in Elliptical Galaxies and Spheroids}",
      journal = {\pasp},
         year = 1998,
        month = aug,
       volume = {110},
       number = {750},
        pages = {888-899},
          doi = {10.1086/316203},
       adsurl = {https://ui.adsabs.harvard.edu/abs/1998PASP..110..888W}
}

@ARTICLE{2000AJ....119.1645T,
       author = {{Trager}, S.~C. and {Faber}, S.~M. and {Worthey}, Guy and {Gonz{\'a}lez}, J. Jes{\'u}s},
        title = "{The Stellar Population Histories of Local Early-Type Galaxies. I. Population Parameters}",
      journal = {\aj},
         year = 2000,
        month = apr,
       volume = {119},
       number = {4},
        pages = {1645-1676},
          doi = {10.1086/301299},
archivePrefix = {arXiv},
       eprint = {astro-ph/0001072},
 primaryClass = {astro-ph},
       adsurl = {https://ui.adsabs.harvard.edu/abs/2000AJ....119.1645T}
}

@ARTICLE{1980FCPh....5..287T,
       author = {{Tinsley}, B.~M.},
        title = "{Evolution of the Stars and Gas in Galaxies}",
      journal = {\fcp},
         year = 1980,
        month = jan,
       volume = {5},
        pages = {287-388},
          doi = {10.48550/arXiv.2203.02041},
archivePrefix = {arXiv},
       eprint = {2203.02041},
 primaryClass = {astro-ph.GA},
       adsurl = {https://ui.adsabs.harvard.edu/abs/1980FCPh....5..287T}
}

@ARTICLE{2013ARA&A..51..393C,
       author = {{Conroy}, Charlie},
        title = "{Modeling the Panchromatic Spectral Energy Distributions of Galaxies}",
      journal = {\araa},
         year = 2013,
        month = aug,
       volume = {51},
       number = {1},
        pages = {393-455},
          doi = {10.1146/annurev-astro-082812-141017},
archivePrefix = {arXiv},
       eprint = {1301.7095},
 primaryClass = {astro-ph.CO},
       adsurl = {https://ui.adsabs.harvard.edu/abs/2013ARA&A..51..393C}
}

@ARTICLE{Tonry1988,
       author = {{Tonry}, John and {Schneider}, Donald P.},
        title = "{A New Technique for Measuring Extragalactic Distances}",
      journal = {\aj},
         year = 1988,
        month = sep,
       volume = {96},
        pages = {807},
          doi = {10.1086/114847},
       adsurl = {https://ui.adsabs.harvard.edu/abs/1988AJ.....96..807T}
}

@ARTICLE{vanDokkum2014,
       author = {{van Dokkum}, Pieter G. and {Conroy}, Charlie},
        title = "{Fluctuation Spectroscopy: A New Probe of Old Stellar Populations}",
      journal = {\apj},
         year = 2014,
        month = dec,
       volume = {797},
       number = {1},
          eid = {56},
        pages = {56},
          doi = {10.1088/0004-637X/797/1/56},
archivePrefix = {arXiv},
       eprint = {1410.0007},
 primaryClass = {astro-ph.GA},
       adsurl = {https://ui.adsabs.harvard.edu/abs/2014ApJ...797...56V}
}

@ARTICLE{Bressan2012,
       author = {{Bressan}, Alessandro and {Marigo}, Paola and {Girardi}, L{\'e}o. and {Salasnich}, Bernardo and {Dal Cero}, Claudia and {Rubele}, Stefano and {Nanni}, Ambra},
        title = "{PARSEC: stellar tracks and isochrones with the PAdova and TRieste Stellar Evolution Code}",
      journal = {\mnras},
         year = 2012,
        month = nov,
       volume = {427},
       number = {1},
        pages = {127-145},
          doi = {10.1111/j.1365-2966.2012.21948.x},
archivePrefix = {arXiv},
       eprint = {1208.4498},
 primaryClass = {astro-ph.SR},
       adsurl = {https://ui.adsabs.harvard.edu/abs/2012MNRAS.427..127B}
}

@ARTICLE{Pastorelli2020,
       author = {{Pastorelli}, Giada and {Marigo}, Paola and {Girardi}, L{\'e}o and {Aringer}, Bernhard and {Chen}, Yang and {Rubele}, Stefano and {Trabucchi}, Michele and {Bladh}, Sara and {Boyer}, Martha L. and {Bressan}, Alessandro and {Dalcanton}, Julianne J. and {Groenewegen}, Martin A.~T. and {Lebzelter}, Thomas and {Mowlavi}, Nami and {Chubb}, Katy L. and {Cioni}, Maria-Rosa L. and {de Grijs}, Richard and {Ivanov}, Valentin D. and {Nanni}, Ambra and {van Loon}, Jacco Th and {Zaggia}, Simone},
        title = "{Constraining the thermally pulsing asymptotic giant branch phase with resolved stellar populations in the Large Magellanic Cloud}",
      journal = {\mnras},
         year = 2020,
        month = nov,
       volume = {498},
       number = {3},
        pages = {3283-3301},
          doi = {10.1093/mnras/staa2565},
archivePrefix = {arXiv},
       eprint = {2008.08595},
 primaryClass = {astro-ph.SR},
       adsurl = {https://ui.adsabs.harvard.edu/abs/2020MNRAS.498.3283P}
}

@ARTICLE{Villaume2017,
       author = {{Villaume}, Alexa and {Conroy}, Charlie and {Johnson}, Benjamin and
         {Rayner}, John and {Mann}, Andrew W. and {van Dokkum}, Pieter},
        title = "{The Extended IRTF Spectral Library: Expanded Coverage in Metallicity, Temperature, and Surface Gravity}",
      journal = {\apjs},
         year = "2017",
        month = "Jun",
       volume = {230},
       number = {2},
          eid = {23},
        pages = {23},
          doi = {10.3847/1538-4365/aa72ed},
archivePrefix = {arXiv},
       eprint = {1705.08906},
 primaryClass = {astro-ph.SR},
       adsurl = {https://ui.adsabs.harvard.edu/abs/2017ApJS..230...23V}
}

@article{Sánchez-Blázquez2006,
    author = {Sánchez-Blázquez, P. and Peletier, R. F. and Jiménez-Vicente, J. and Cardiel, N. and Cenarro, A. J. and Falcón-Barroso, J. and Gorgas, J. and Selam, S. and Vazdekis, A.},
    title = {Medium-resolution Isaac Newton Telescope library of empirical spectra},
    journal = {Monthly Notices of the Royal Astronomical Society},
    volume = {371},
    number = {2},
    pages = {703-718},
    year = {2006},
    month = {08},
    issn = {0035-8711},
    doi = {10.1111/j.1365-2966.2006.10699.x},
    url = {https://doi.org/10.1111/j.1365-2966.2006.10699.x},
    eprint = {https://academic.oup.com/mnras/article-pdf/371/2/703/3443797/mnras0371-0703.pdf},
}

@ARTICLE{Salpeter1955,
       author = {{Salpeter}, Edwin E.},
        title = "{The Luminosity Function and Stellar Evolution.}",
      journal = {\apj},
         year = 1955,
        month = jan,
       volume = {121},
        pages = {161},
          doi = {10.1086/145971},
       adsurl = {https://ui.adsabs.harvard.edu/abs/1955ApJ...121..161S}
}

@ARTICLE{Elham2022,
       author = {{Eftekhari}, Elham and {La Barbera}, Francesco and {Vazdekis}, Alexandre and {Allende Prieto}, Carlos and {Knowles}, Adam Thomas},
        title = "{Strong CO absorption features in massive ETGs}",
      journal = {\mnras},
         year = 2022,
        month = may,
       volume = {512},
       number = {1},
        pages = {378-400},
          doi = {10.1093/mnras/stac471},
archivePrefix = {arXiv},
       eprint = {2202.08651},
 primaryClass = {astro-ph.GA},
       adsurl = {https://ui.adsabs.harvard.edu/abs/2022MNRAS.512..378E}
}

@ARTICLE{Baldwin2018,
       author = {{Baldwin}, C. and {McDermid}, R.~M. and {Kuntschner}, H. and {Maraston}, C. and {Conroy}, C.},
        title = "{Comparison of stellar population model predictions using optical and infrared spectroscopy}",
      journal = {\mnras},
         year = 2018,
        month = feb,
       volume = {473},
       number = {4},
        pages = {4698-4721},
          doi = {10.1093/mnras/stx2502},
archivePrefix = {arXiv},
       eprint = {1709.09300},
 primaryClass = {astro-ph.GA},
       adsurl = {https://ui.adsabs.harvard.edu/abs/2018MNRAS.473.4698B}
}

@ARTICLE{Beattie2021,
       author = {{Beattie}, J. Renwick and {Esmonde-White}, Francis W.~L.},
        title = "{Exploration of Principal Component Analysis: Deriving Principal Component Analysis Visually Using Spectra}",
      journal = {Applied Spectroscopy},
         year = 2021,
        month = apr,
       volume = {75},
       number = {4},
        pages = {361-375},
          doi = {10.1177/0003702820987847},
       adsurl = {https://ui.adsabs.harvard.edu/abs/2021ApSpe..75..361B}
}

@article{pearson1901,
  author       = {Pearson, Karl},
  title        = {On lines and planes of closest fit to systems of points in space},
  journal      = {The London, Edinburgh, and Dublin Philosophical Magazine and Journal of Science},
  volume       = {2},
  number       = {11},
  pages        = {559--572},
  year         = {1901},
  doi          = {10.1080/14786440109462720},
  publisher    = {Taylor \& Francis}
}

@article{jolliffe2016,
  author       = {Jolliffe, Ian T. and Cadima, Jorge},
  title        = {Principal component analysis: a review and recent developments},
  journal      = {Philosophical Transactions of the Royal Society A: Mathematical, Physical and Engineering Sciences},
  volume       = {374},
  number       = {2065},
  pages        = {20150202},
  year         = {2016},
  doi          = {10.1098/rsta.2015.0202},
}

@ARTICLE{LaBarbera2025,
       author = {{La Barbera}, F. and {Vazdekis}, A. and {Pasquali}, A. and {Heidt}, J. and {Eftekhari}, E. and {Beasley}, M.~A. and {Gargiulo}, A. and {Bisogni}, S. and {Spiniello}, C. and {Cassar{\`a}}, L.~P. and {Sarzi}, M.},
        title = "{Puzzling radial gradients of K-band absorption features in the giant elliptical galaxy M87}",
      journal = {\aap},
         year = 2025,
        month = aug,
       volume = {700},
          eid = {A64},
        pages = {A64},
          doi = {10.1051/0004-6361/202554315},
archivePrefix = {arXiv},
       eprint = {2506.05986},
 primaryClass = {astro-ph.GA},
       adsurl = {https://ui.adsabs.harvard.edu/abs/2025A&A...700A..64L}
}

@ARTICLE{Marigo2017,
       author = {{Marigo}, Paola and {Girardi}, L{\'e}o and {Bressan}, Alessandro and {Rosenfield}, Philip and {Aringer}, Bernhard and {Chen}, Yang and {Dussin}, Marco and {Nanni}, Ambra and {Pastorelli}, Giada and {Rodrigues}, Tha{\'\i}se S. and {Trabucchi}, Michele and {Bladh}, Sara and {Dalcanton}, Julianne and {Groenewegen}, Martin A.~T. and {Montalb{\'a}n}, Josefina and {Wood}, Peter R.},
        title = "{A New Generation of PARSEC-COLIBRI Stellar Isochrones Including the TP-AGB Phase}",
      journal = {\apj},
         year = 2017,
        month = jan,
       volume = {835},
       number = {1},
          eid = {77},
        pages = {77},
          doi = {10.3847/1538-4357/835/1/77},
archivePrefix = {arXiv},
       eprint = {1701.08510},
 primaryClass = {astro-ph.SR},
       adsurl = {https://ui.adsabs.harvard.edu/abs/2017ApJ...835...77M}
}

@ARTICLE{Toft2014,
       author = {{Toft}, S. and {Smol{\v{c}}i{\'c}}, V. and {Magnelli}, B. and {Karim}, A. and {Zirm}, A. and {Michalowski}, M. and {Capak}, P. and {Sheth}, K. and {Schawinski}, K. and {Krogager}, J.-K. and {Wuyts}, S. and {Sanders}, D. and {Man}, A.~W.~S. and {Lutz}, D. and {Staguhn}, J. and {Berta}, S. and {Mccracken}, H. and {Krpan}, J. and {Riechers}, D.},
        title = "{Submillimeter Galaxies as Progenitors of Compact Quiescent Galaxies}",
      journal = {\apj},
         year = 2014,
        month = feb,
       volume = {782},
       number = {2},
          eid = {68},
        pages = {68},
          doi = {10.1088/0004-637X/782/2/68},
archivePrefix = {arXiv},
       eprint = {1401.1510},
 primaryClass = {astro-ph.GA},
       adsurl = {https://ui.adsabs.harvard.edu/abs/2014ApJ...782...68T}
}

@ARTICLE{Costa2024,
       author = {{Costa}, G. and {Dumont}, T. and {Lan{\c{c}}on}, A. and {Palacios}, A. and {Charbonnel}, C. and {Prugniel}, P. and {Ekstrom}, S. and {Georgy}, C. and {Branco}, V. and {Coelho}, P. and {Martins}, L. and {Borisov}, S. and {Voggel}, K. and {Chantereau}, W.},
        title = "{He-enriched STAREVOL models for globular cluster multiple populations: Self-consistent isochrones from ZAMS to the TP-AGB phase}",
      journal = {\aap},
         year = 2024,
        month = oct,
       volume = {690},
          eid = {A22},
        pages = {A22},
          doi = {10.1051/0004-6361/202449741},
archivePrefix = {arXiv},
       eprint = {2405.16505},
 primaryClass = {astro-ph.SR},
       adsurl = {https://ui.adsabs.harvard.edu/abs/2024A&A...690A..22C}
}

@ARTICLE{Cameron2023,
       author = {{Cameron}, Alex J. and {Katz}, Harley and {Rey}, Martin P. and {Saxena}, Aayush},
        title = "{Nitrogen enhancements 440 Myr after the big bang: supersolar N/O, a tidal disruption event, or a dense stellar cluster in GN-z11?}",
      journal = {\mnras},
         year = 2023,
        month = aug,
       volume = {523},
       number = {3},
        pages = {3516-3525},
          doi = {10.1093/mnras/stad1579},
archivePrefix = {arXiv},
       eprint = {2302.10142},
 primaryClass = {astro-ph.GA},
       adsurl = {https://ui.adsabs.harvard.edu/abs/2023MNRAS.523.3516C}
}

@ARTICLE{Chung2017,
       author = {{Chung}, Chul and {Yoon}, Suk-Jin and {Lee}, Young-Wook},
        title = "{Yonsei Evolutionary Population Synthesis (YEPS). II. Spectro-photometric Evolution of Helium-enhanced Stellar Populations}",
      journal = {\apj},
         year = 2017,
        month = jun,
       volume = {842},
       number = {2},
          eid = {91},
        pages = {91},
          doi = {10.3847/1538-4357/aa6f19},
archivePrefix = {arXiv},
       eprint = {1704.07382},
 primaryClass = {astro-ph.GA},
       adsurl = {https://ui.adsabs.harvard.edu/abs/2017ApJ...842...91C}
}

@INPROCEEDINGS{Ali2023,
       author = {{Ali}, Sadman S.},
        title = "{Origin of the excess Ultraviolet light in Early-type Galaxies: A case of Helium-enhanced stellar populations at late stages of evolution}",
    booktitle = {American Astronomical Society Meeting Abstracts \#241},
         year = 2023,
       series = {American Astronomical Society Meeting Abstracts},
       volume = {241},
        month = jan,
          eid = {319.06},
        pages = {319.06},
       adsurl = {https://ui.adsabs.harvard.edu/abs/2023AAS...24131906A}
}

@ARTICLE{Gilroy1991,
       author = {{Gilroy}, Kalpana Krishnaswamy and {Brown}, Jeffery A.},
        title = "{Carbon Isotope Ratios along the Giant Branch of M67}",
      journal = {\apj},
         year = 1991,
        month = apr,
       volume = {371},
        pages = {578},
          doi = {10.1086/169922},
       adsurl = {https://ui.adsabs.harvard.edu/abs/1991ApJ...371..578G}
}

@ARTICLE{2020arXiv200909242M,
       author = {{McDermid}, Richard M. and {Cresci}, Giovanni and {Rigaut}, Francois and {Bouret}, Jean-Claude and {De Silva}, Gayandhi and {Gullieuszik}, Marco and {Magrini}, Laura and {Mendel}, J. Trevor and {Antoniucci}, Simone and {Bono}, Giuseppe and {Kamath}, Devika and {Monty}, Stephanie and {Baumgardt}, Holger and {Cortese}, Luca and {Fisher}, Deanne and {Mannucci}, Filippo and {Migliorini}, Alessandra and {Sweet}, Sarah and {Vanzella}, Eros and {Zibetti}, Stefano and {White Papers.}, with additional contributions from the authors of the MAVIS},
        title = "{Phase A Science Case for MAVIS -- The Multi-conjugate Adaptive-optics Visible Imager-Spectrograph for the VLT Adaptive Optics Facility}",
      journal = {arXiv e-prints},
         year = 2020,
        month = sep,
          eid = {arXiv:2009.09242},
        pages = {arXiv:2009.09242},
          doi = {10.48550/arXiv.2009.09242},
archivePrefix = {arXiv},
       eprint = {2009.09242},
 primaryClass = {astro-ph.IM},
       adsurl = {https://ui.adsabs.harvard.edu/abs/2020arXiv200909242M}
}

@ARTICLE{2023A&A...674A.207D,
       author = {{Davies}, R. and {Absil}, O. and {Agapito}, G. and {Agudo Berbel}, A. and {Baruffolo}, A. and {Biliotti}, V. and {Black}, M. and {Bonaglia}, M. and {Bonse}, M. and {Briguglio}, R. and {Campana}, P. and {Cao}, Y. and {Carbonaro}, L. and {Cortes}, A. and {Cresci}, G. and {Dallilar}, Y. and {Dannert}, F. and {De Rosa}, R.~J. and {Deysenroth}, M. and {Di Antonio}, I. and {Di Cianno}, A. and {Di Rico}, G. and {Doelman}, D. and {Dolci}, M. and {Dorn}, R. and {Eisenhauer}, F. and {Esposito}, S. and {Fantinel}, D. and {Ferruzzi}, D. and {Feuchtgruber}, H. and {Finger}, G. and {F{\"o}rster Schreiber}, N.~M. and {Gao}, X. and {Gemperlein}, H. and {Genzel}, R. and {Gillessen}, S. and {Ginski}, C. and {Glauser}, A.~M. and {Glindemann}, A. and {Grani}, P. and {Hartl}, M. and {Hayoz}, J. and {Heida}, M. and {Henry}, D. and {Hofmann}, R. and {Huber}, H. and {Kasper}, M. and {Keller}, C. and {Kenworthy}, M. and {Kravchenko}, K. and {Kuntschner}, H. and {Lacour}, S. and {Lightfoot}, J. and {Lunney}, D. and {Lutz}, D. and {Macintosh}, M. and {Mannucci}, F. and {Marsset}, M. and {Modigliani}, A. and {Neeser}, M. and {Orban de Xivry}, G. and {Ott}, T. and {Pallanca}, L. and {Patapis}, P. and {Pearson}, D. and {Pe{\~n}a}, E. and {Percheron}, I. and {Puglisi}, A. and {Quanz}, S.~P. and {Rabien}, S. and {Rau}, C. and {Riccardi}, A. and {Salasnich}, B. and {Schmid}, H.-M. and {Schubert}, J. and {Serra}, B. and {Shimizu}, T. and {Snik}, F. and {Sturm}, E. and {Tacconi}, L. and {Taylor}, W. and {Valentini}, A. and {Waring}, C. and {Wiezorrek}, E. and {Xompero}, M.},
        title = "{The Enhanced Resolution Imager and Spectrograph for the VLT}",
      journal = {\aap},
         year = 2023,
        month = jun,
       volume = {674},
          eid = {A207},
        pages = {A207},
          doi = {10.1051/0004-6361/202346559},
archivePrefix = {arXiv},
       eprint = {2304.02343},
 primaryClass = {astro-ph.IM},
       adsurl = {https://ui.adsabs.harvard.edu/abs/2023A&A...674A.207D}
}

@ARTICLE{Buzzoni1989,
       author = {{Buzzoni}, Alberto},
        title = "{Evolutionary Population Synthesis in Stellar Systems. I. A Global Approach}",
      journal = {\apjs},
         year = 1989,
        month = dec,
       volume = {71},
        pages = {817},
          doi = {10.1086/191399},
       adsurl = {https://ui.adsabs.harvard.edu/abs/1989ApJS...71..817B}
}

@ARTICLE{Buzzoni2005,
       author = {{Buzzoni}, Alberto},
        title = "{Photometric entropy of stellar populations and related diagnostic tools}",
      journal = {arXiv e-prints},
         year = 2005,
        month = sep,
          eid = {astro-ph/0509602},
        pages = {astro-ph/0509602},
          doi = {10.48550/arXiv.astro-ph/0509602},
archivePrefix = {arXiv},
       eprint = {astro-ph/0509602},
 primaryClass = {astro-ph},
       adsurl = {https://ui.adsabs.harvard.edu/abs/2005astro.ph..9602B}
}

@ARTICLE{Cervino2006,
       author = {{Cervi{\~n}o}, M. and {Luridiana}, V.},
        title = "{Confidence limits of evolutionary synthesis models. IV. Moving forward to a probabilistic formulation}",
      journal = {\aap},
         year = 2006,
        month = may,
       volume = {451},
       number = {2},
        pages = {475-498},
          doi = {10.1051/0004-6361:20053283},
archivePrefix = {arXiv},
       eprint = {astro-ph/0504483},
 primaryClass = {astro-ph},
       adsurl = {https://ui.adsabs.harvard.edu/abs/2006A&A...451..475C}
}

@ARTICLE{Tinsley1980,
       author = {{Tinsley}, B.~M.},
        title = "{Evolution of the Stars and Gas in Galaxies}",
      journal = {\fcp},
         year = 1980,
        month = jan,
       volume = {5},
        pages = {287-388},
          doi = {10.48550/arXiv.2203.02041},
archivePrefix = {arXiv},
       eprint = {2203.02041},
 primaryClass = {astro-ph.GA},
       adsurl = {https://ui.adsabs.harvard.edu/abs/1980FCPh....5..287T}
}

@ARTICLE{1993A&A...275..433B,
       author = {{Buzzoni}, A.},
        title = "{Statistical properties of stellar populations and surface-brightness fluctuations in galaxies.}",
      journal = {\aap},
         year = 1993,
        month = aug,
       volume = {275},
        pages = {433-450},
       adsurl = {https://ui.adsabs.harvard.edu/abs/1993A&A...275..433B}
}

@ARTICLE{2020MNRAS.493.5131V,
       author = {{Vazdekis}, A. and {Cervi{\~n}o}, M. and {Montes}, M. and {Mart{\'\i}n-Navarro}, I. and {Beasley}, M.~A.},
        title = "{Surface brightness fluctuation spectra to constrain stellar population properties}",
      journal = {\mnras},
         year = 2020,
        month = apr,
       volume = {493},
       number = {4},
        pages = {5131-5152},
          doi = {10.1093/mnras/staa629},
archivePrefix = {arXiv},
       eprint = {2003.02563},
 primaryClass = {astro-ph.GA},
       adsurl = {https://ui.adsabs.harvard.edu/abs/2020MNRAS.493.5131V}
}

@ARTICLE{1993ApJ...409..530W,
       author = {{Worthey}, Guy},
        title = "{The Dependence of the Brightness Fluctuation Distance Indicator on Stellar Population Age and Metallicity}",
      journal = {\apj},
         year = 1993,
        month = jun,
       volume = {409},
        pages = {530},
          doi = {10.1086/172684},
       adsurl = {https://ui.adsabs.harvard.edu/abs/1993ApJ...409..530W}
}

@ARTICLE{2007ApJ...668..130C,
       author = {{Cantiello}, Michele and {Blakeslee}, John and {Raimondo}, Gabriella and {Brocato}, Enzo and {Capaccioli}, Massimo},
        title = "{Surface Brightness Fluctuations from Archival ACS Images: A Stellar Population and Distance Study}",
      journal = {\apj},
         year = 2007,
        month = oct,
       volume = {668},
       number = {1},
        pages = {130-149},
          doi = {10.1086/521295},
archivePrefix = {arXiv},
       eprint = {0706.4467},
 primaryClass = {astro-ph},
       adsurl = {https://ui.adsabs.harvard.edu/abs/2007ApJ...668..130C}
}

@ARTICLE{2021MNRAS.507.3005R,
       author = {{Rodr{\'\i}guez-Beltr{\'a}n}, P. and {Vazdekis}, A. and {Cervi{\~n}o}, M. and {Beasley}, M.~A.},
        title = "{Surface brightness fluctuations to constrain secondary stellar populations: revealing very low-metallicity stars in massive galaxies}",
      journal = {\mnras},
         year = 2021,
        month = oct,
       volume = {507},
       number = {2},
        pages = {3005-3029},
          doi = {10.1093/mnras/stab2330},
archivePrefix = {arXiv},
       eprint = {2107.08722},
 primaryClass = {astro-ph.GA},
       adsurl = {https://ui.adsabs.harvard.edu/abs/2021MNRAS.507.3005R}
}



\appendix

\section{Fractional variance contributions for all the synthesised models}
\label{App-table}

Table\,\ref{tab:full-stat-table} lists the full set of cases explored including the SSP and mixed parameter spaces. The SSP cases cover combinations of three metallicities used (-- 0.3, 0.0, and + 0.3 dex) with the four explored ages (3, 6, 10, and 13 Gyr). The mixed cases incorporate the extreme cases of iron abundance (-- 0.3 \& + 0.3 dex) comiled for all ages, likewise, the mixed ages (3 \& 13 Gyr) were computed for all explored metallicities.
Two isochrones are combined in each realisation according to their total luminosity, so that both components contribute equally to the integrated light. Each pixel receives an even 50:50 contribution from the combined cases. The total luminosity used to scale the realisations is the same in both the combined SSP and single SSP cases.

Similarly to the Table\,\ref{tab:basecases-stat} the fractional variance contributions of the first four components in the PCA decomposition are listed, along with the quantities $\bar{L} = {\rm var}(L)/\langle L\rangle$, corresponding to the classic SBF luminosity (i.e. the characteristic luminosity of the stars driving the fluctuations), and $\bar{N} = \langle L\rangle/\bar{L} = \langle L\rangle^2/{\rm var}(L)$, which represents the effective number of stars per sample.
As expected, fluctuations in the redder bands are dominated by only a few very bright red giants in each synthesised spectrum. Across all bands, with PC1 and PC2 together accounting for over 95 per cent of the total variance, which confirms that using only the first two components is sufficient.

\onecolumn

\begin{longtable}{cc|cccc|cc}

\caption{Complete statistical table for all performed parameter space.} \\
\label{tab:full-stat-table}

Fe/H & Age & PC1 & PC2 & PC3 & PC4  &  $\bar{L}/L_\odot$ & 
$\bar{N}$ \\
\endfirsthead

\hline
Fe/H & Age & PC1 & PC2 & PC3 & PC4  &  $\bar{L}/L_\odot$ & 
$\bar{N}$ \\
\hline
\endhead

\hline
\endfoot

 \hline \multicolumn{8}{c}{V-band} \\ \hline
+ 0.0 & 3 & \bf{87.19} & \bf{11.53} & 0.89 & 0.27 & 38.17 & 261.81 \\
+ 0.0 & 6 & \bf{87.19} & \bf{11.38} & 0.94 & 0.32 & 29.61 & 338.90 \\
+ 0.0 & 10 & \bf{89.69} & \bf{9.02} & 0.78 & 0.38 & 27.89 & 358.67 \\
+ 0.0 & 13 & \bf{90.31} & \bf{8.08} & 1.00 & 0.43 & 23.93 & 417.54 \\
+ 0.3 & 3 & \bf{79.41} & \bf{18.21} & 1.78 & 0.35 & 25.40 & 393.07 \\
+ 0.3 & 6 & \bf{83.35} & \bf{14.73} & 1.28 & 0.38 & 20.23 & 493.52 \\
+ 0.3 & 10 & \bf{87.23} & \bf{10.70} & 1.39 & 0.42 & 15.90 & 628.85 \\
+ 0.3 & 13 & \bf{86.24} & \bf{11.09} & 1.85 & 0.55 & 17.50 & 571.10 \\
-- 0.3 & 3 & \bf{88.52} & \bf{10.19} & 0.88 & 0.28 & 71.77 & 139.02 \\
-- 0.3 & 6 & \bf{89.65} & \bf{9.27} & 0.71 & 0.23 & 50.11 & 199.82 \\
-- 0.3 & 10 & \bf{91.38} & \bf{7.63} & 0.60 & 0.28 & 46.43 & 216.34 \\
-- 0.3 & 13 & \bf{92.75} & \bf{6.17} & 0.58 & 0.36 & 42.54 & 235.43 \\

\hline \multicolumn{8}{c}{V-band mixed} \\ \hline
-- 0.3 \& + 0.3 & 3 & \bf{87.00} & \bf{10.65} & 1.02 & 0.71 & 46.97 & 212.09 \\
-- 0.3 \& + 0.3 & 6 & \bf{87.72} & \bf{9.10} & 1.63 & 0.75 & 34.67 & 287.78 \\
-- 0.3 \& + 0.3 & 10 & \bf{89.93} & \bf{8.08} & 0.82 & 0.65 & 29.26 & 342.32 \\
-- 0.3 \& + 0.3 & 13 & \bf{91.86} & \bf{6.31} & 0.78 & 0.56 & 28.82 & 347.82 \\
+ 0.0 & 3 \& 13 & \bf{87.54} & \bf{11.02} & 0.84 & 0.37 & 32.58 & 306.28 \\
+ 0.3 & 3 \& 13 & \bf{83.40} & \bf{14.06} & 1.68 & 0.44 & 20.05 & 498.71 \\
-- 0.3 & 3 \& 13 & \bf{90.68} & \bf{8.17} & 0.69 & 0.29 & 55.51 & 179.84 \\

\hline \multicolumn{8}{c}{I-band} \\ \hline
+ 0.0 & 3 & \bf{86.31} & \bf{12.23} & 1.25 & 0.15 & 229.28 & 43.73 \\
+ 0.0 & 6 & \bf{90.05} & \bf{8.53} & 1.21 & 0.16 & 163.91 & 61.25 \\
+ 0.0 & 10 & \bf{92.20} & \bf{6.83} & 0.83 & 0.10 & 129.41 & 77.68 \\
+ 0.0 & 13 & \bf{92.14} & \bf{6.98} & 0.76 & 0.08 & 126.96 & 78.91 \\
+ 0.3 & 3 & \bf{83.03} & \bf{14.69} & 1.94 & 0.27 & 146.60 & 68.72 \\
+ 0.3 & 6 & \bf{88.68} & \bf{9.73} & 1.32 & 0.20 & 94.39 & 106.08 \\
+ 0.3 & 10 & \bf{86.93} & \bf{11.46} & 1.35 & 0.20 & 81.84 & 122.30 \\
+ 0.3 & 13 & \bf{86.96} & \bf{11.45} & 1.33 & 0.19 & 74.59 & 133.69 \\
-- 0.3 & 3 & \bf{89.00} & \bf{9.72} & 1.12 & 0.12 & 284.47 & 34.81 \\
-- 0.3 & 6 & \bf{92.10} & \bf{6.72} & 1.06 & 0.08 & 235.08 & 42.70 \\
-- 0.3 & 10 & \bf{95.45} & 3.85 & 0.62 & 0.05 & 198.58 & 50.15 \\
-- 0.3 & 13 & \bf{96.04} & 3.38 & 0.50 & 0.04 & 182.52 & 54.82 \\

\hline \multicolumn{8}{c}{I-band mixed} \\ \hline
-- 0.3 \& + 0.3 & 3 & \bf{86.62} & \bf{11.38} & 1.38 & 0.28 & 231.26 & 43.41 \\
-- 0.3 \& + 0.3 & 6 & \bf{87.06} & \bf{10.86} & 1.31 & 0.43 & 169.82 & 58.80 \\
-- 0.3 \& + 0.3 & 10 & \bf{91.82} & \bf{6.68} & 1.05 & 0.20 & 156.44 & 64.25 \\
-- 0.3 \& + 0.3 & 13 & \bf{93.41} & \bf{5.45} & 0.75 & 0.19 & 135.47 & 73.67 \\
+ 0.0 & 3 \& 13 & \bf{88.51} & \bf{10.25} & 1.04 & 0.14 & 183.69 & 54.55 \\
+ 0.3 & 3 \& 13 & \bf{84.40} & \bf{13.70} & 1.60 & 0.23 & 109.85 & 91.25 \\
-- 0.3 & 3 \& 13 & \bf{90.49} & \bf{8.34} & 1.01 & 0.11 & 246.78 & 40.50 \\

\hline \multicolumn{8}{c}{J-band} \\ \hline
+ 0.0 & 3 & \bf{83.04} & \bf{13.01} & 3.08 & 0.75 & 1880.14 & 5.37 \\
+ 0.0 & 6 & \bf{76.13} & \bf{19.15} & 3.70 & 0.89 & 1226.81 & 8.13 \\
+ 0.0 & 10 & \bf{78.80} & \bf{17.60} & 2.77 & 0.68 & 901.39 & 11.12 \\
+ 0.0 & 13 & \bf{81.98} & \bf{15.53} & 2.09 & 0.26 & 840.19 & 11.96 \\
+ 0.3 & 3 & \bf{85.07} & \bf{10.90} & 2.95 & 0.78 & 1538.65 & 6.46 \\
+ 0.3 & 6 & \bf{85.95} & \bf{10.50} & 2.60 & 0.74 & 1080.77 & 9.20 \\
+ 0.3 & 10 & \bf{88.42} & \bf{8.53} & 2.22 & 0.64 & 1014.46 & 9.84 \\
+ 0.3 & 13 & \bf{88.64} & \bf{8.42} & 1.84 & 0.83 & 1012.66 & 9.82 \\
-- 0.3 & 3 & \bf{63.22} & \bf{33.06} & 3.14 & 0.37 & 1429.11 & 7.13 \\
-- 0.3 & 6 & \bf{68.29} & \bf{27.53} & 3.82 & 0.22 & 793.61 & 12.59 \\
-- 0.3 & 10 & \bf{84.56} & \bf{12.85} & 2.29 & 0.22 & 652.34 & 15.26 \\
-- 0.3 & 13 & \bf{79.23} & \bf{17.77} & 2.59 & 0.25 & 636.50 & 15.61 \\

\hline \multicolumn{8}{c}{J-band mixed} \\ \hline
-- 0.3 \& + 0.3 & 3 & \bf{74.23} & \bf{18.51} & 4.86 & 1.16 & 1664.52 & 6.09 \\
-- 0.3 \& + 0.3 & 6 & \bf{77.80} & \bf{15.86} & 3.91 & 1.64 & 978.44 & 10.27 \\
-- 0.3 \& + 0.3 & 10 & \bf{66.43} & \bf{26.64} & 4.07 & 2.16 & 878.67 & 11.43 \\
-- 0.3 \& + 0.3 & 13 & \bf{65.82} & \bf{27.43} & 3.86 & 2.02 & 818.67 & 12.37 \\
+ 0.0 & 3 \& 13 & \bf{76.38} & \bf{18.79} & 3.67 & 0.99 & 1159.00 & 8.50 \\
+ 0.3 & 3 \& 13 & \bf{86.52} & \bf{9.88} & 2.47 & 0.78 & 1298.67 & 7.86 \\
-- 0.3 & 3 \& 13 & \bf{60.22} & \bf{35.34} & 3.76 & 0.44 & 1113.01 & 9.04 \\

\hline \multicolumn{8}{c}{H-band} \\ \hline
+ 0.0 & 3 & \bf{88.85} & \bf{9.70} & 1.26 & 0.10 & 3972.83 & 2.49 \\
+ 0.0 & 6 & \bf{86.60} & \bf{12.13} & 1.08 & 0.10 & 2514.80 & 3.98 \\
+ 0.0 & 10 & \bf{92.55} & \bf{6.33} & 0.95 & 0.11 & 1780.64 & 5.48 \\
+ 0.0 & 13 & \bf{95.02} & 4.17 & 0.66 & 0.10 & 1544.30 & 6.42 \\
+ 0.3 & 3 & \bf{85.66} & \bf{12.64} & 1.10 & 0.46 & 3111.52 & 3.19 \\
+ 0.3 & 6 & \bf{87.64} & \bf{11.00} & 1.02 & 0.19 & 2401.77 & 4.28 \\
+ 0.3 & 10 & \bf{89.47} & \bf{9.38} & 0.84 & 0.18 & 2140.50 & 4.60 \\
+ 0.3 & 13 & \bf{87.14} & \bf{9.45} & 2.78 & 0.48 & 1863.44 & 5.35 \\
-- 0.3 & 3 & \bf{90.49} & \bf{7.94} & 1.37 & 0.09 & 3524.05 & 2.98 \\
-- 0.3 & 6 & \bf{93.86} & 4.69 & 1.26 & 0.13 & 1571.44 & 6.22 \\
-- 0.3 & 10 & \bf{96.74} & 2.68 & 0.38 & 0.14 & 1397.21 & 7.20 \\
-- 0.3 & 13 & \bf{96.49} & 2.79 & 0.49 & 0.15 & 1592.59 & 6.28 \\

\hline \multicolumn{8}{c}{H-band mixed} \\ \hline
-- 0.3 \& + 0.3 & 3 & \bf{87.52} & \bf{10.34} & 1.11 & 0.70 & 3380.62 & 2.97 \\
-- 0.3 \& + 0.3 & 6 & \bf{83.06} & \bf{14.68} & 1.23 & 0.65 & 2093.12 & 4.82 \\
-- 0.3 \& + 0.3 & 10 & \bf{79.78} & \bf{17.35} & 1.38 & 1.03 & 1489.30 & 6.57 \\
-- 0.3 \& + 0.3 & 13 & \bf{81.30} & \bf{15.27} & 1.42 & 0.98 & 1716.55 & 5.92 \\
+ 0.0 & 3 \& 13 & \bf{81.64} & \bf{17.28} & 0.89 & 0.10 & 2596.99 & 3.88 \\
+ 0.3 & 3 \& 13 & \bf{92.00} & \bf{6.47} & 1.27 & 0.19 & 2452.89 & 4.08 \\
-- 0.3 & 3 \& 13 & \bf{84.55} & \bf{13.58} & 1.65 & 0.10 & 2498.07 & 4.08 \\

\hline \multicolumn{8}{c}{K-band} \\ \hline
+ 0.0 & 3 & \bf{93.17} & \bf{6.49} & 0.26 & 0.06 & 5395.43 & 1.79 \\
+ 0.0 & 6 & \bf{93.47} & \bf{6.21} & 0.24 & 0.05 & 3739.48 & 2.70 \\
+ 0.0 & 10 & \bf{95.55} & 4.18 & 0.21 & 0.04 & 2789.79 & 3.57 \\
+ 0.0 & 13 & \bf{96.89} & 2.90 & 0.15 & 0.04 & 2568.63 & 4.02 \\
+ 0.3 & 3 & \bf{93.07} & \bf{6.53} & 0.27 & 0.08 & 4997.91 & 1.92 \\
+ 0.3 & 6 & \bf{94.00} & \bf{5.66} & 0.24 & 0.05 & 3922.51 & 2.55 \\
+ 0.3 & 10 & \bf{94.23} & \bf{5.41} & 0.26 & 0.05 & 3309.60 & 2.92 \\
+ 0.3 & 13 & \bf{95.13} & 4.59 & 0.19 & 0.04 & 3476.16 & 2.92 \\
-- 0.3 & 3 & \bf{94.95} & 4.58 & 0.34 & 0.11 & 4901.94 & 2.08 \\
-- 0.3 & 6 & \bf{96.70} & 2.90 & 0.29 & 0.10 & 2443.51 & 4.12 \\
-- 0.3 & 10 & \bf{98.68} & 1.02 & 0.17 & 0.11 & 2175.81 & 4.60 \\
-- 0.3 & 13 & \bf{98.36} & 1.30 & 0.21 & 0.12 & 1738.84 & 5.80 \\

\hline \multicolumn{8}{c}{K-band mixed} \\ \hline
-- 0.3 \& + 0.3 & 3 & \bf{93.32} & \bf{5.51} & 0.82 & 0.21 & 5726.68 & 1.80 \\
-- 0.3 \& + 0.3 & 6 & \bf{93.82} & 4.90 & 0.84 & 0.24 & 3308.53 & 3.05 \\
-- 0.3 \& + 0.3 & 10 & \bf{92.87} & \bf{6.03} & 0.70 & 0.25 & 2619.00 & 3.78 \\
-- 0.3 \& + 0.3 & 13 & \bf{93.46} & \bf{5.24} & 0.74 & 0.32 & 2228.79 & 4.40 \\
+ 0.0 & 3 \& 13 & \bf{93.57} & \bf{6.06} & 0.28 & 0.06 & 3891.49 & 2.51 \\
+ 0.3 & 3 \& 13 & \bf{93.33} & \bf{6.19} & 0.32 & 0.10 & 4757.19 & 2.11 \\
-- 0.3 & 3 \& 13 & \bf{95.93} & 3.61 & 0.33 & 0.11 & 3357.69 & 3.01 \\

\end{longtable}
\twocolumn

\section{SSP-like comparisons for V, J, H, and K band }
\label{apx:ages_comparison_bands}
\begin{figure}
 \includegraphics[width=\columnwidth]{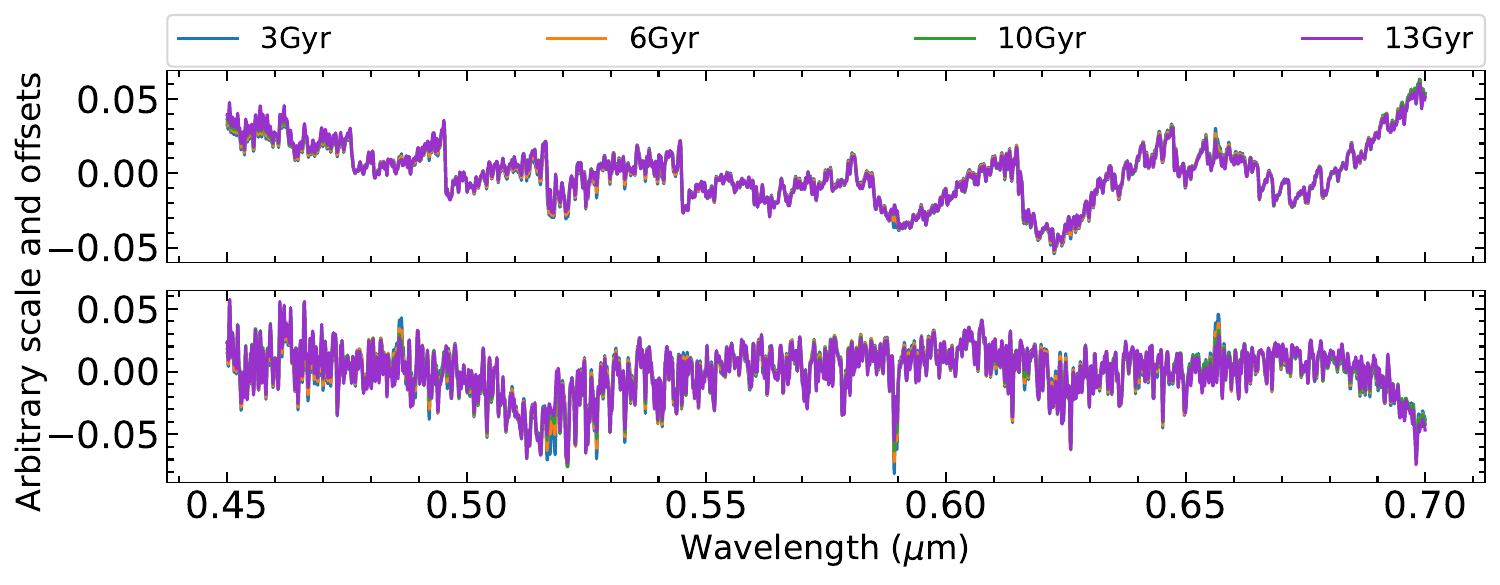}
 \caption{Same as Figure\,\ref{fig:age-variation_Iband} for the V-band.}
 \label{fig:age-variation_Vband}
\end{figure}

\begin{figure}
 \includegraphics[width=\columnwidth]{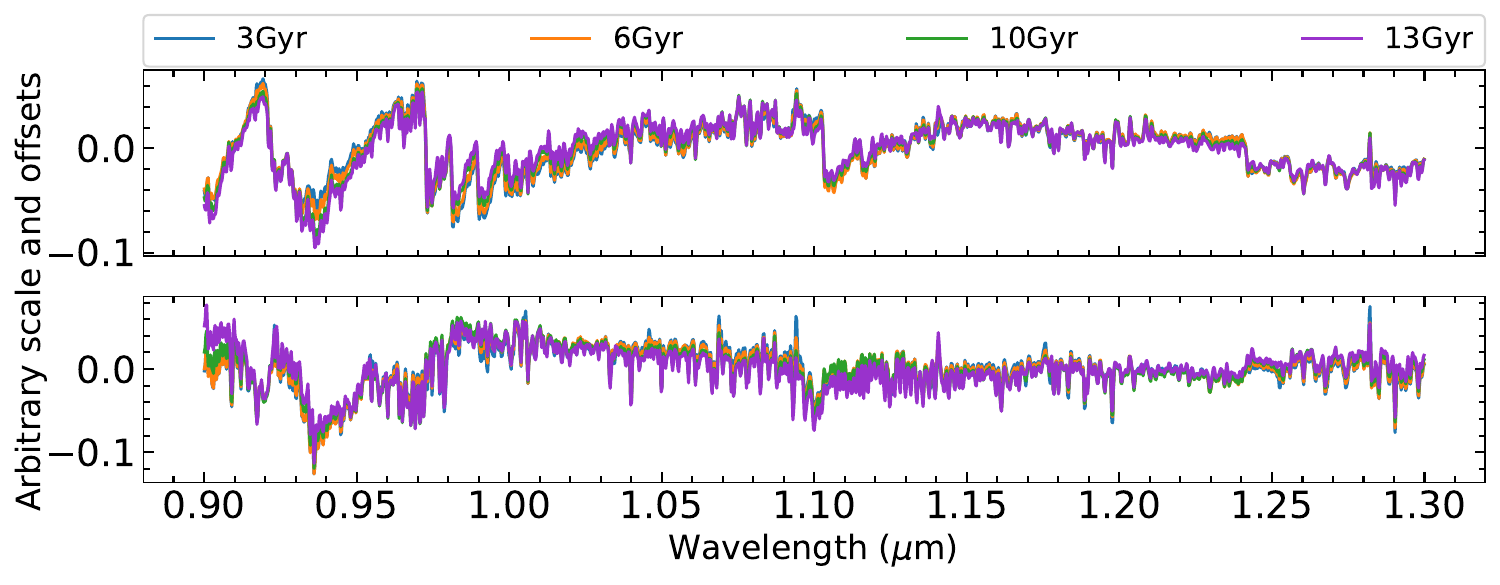}
 \caption{Same as Figure\,\ref{fig:age-variation_Iband} for the J-band.}
 \label{fig:age-variation_Jband}
\end{figure}

\begin{figure}
 \includegraphics[width=\columnwidth]{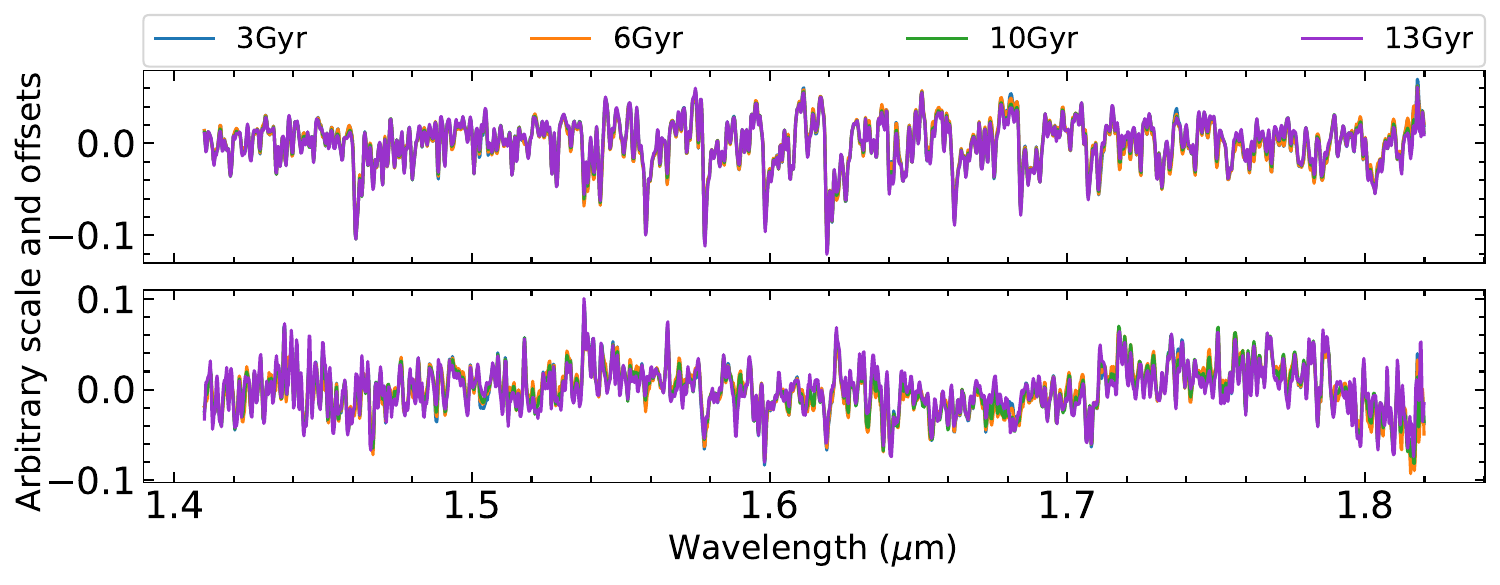}
 \caption{Same as Figure\,\ref{fig:age-variation_Iband} for the H-band.}
 \label{fig:age-variation_Hband}
\end{figure}

\begin{figure}
 \includegraphics[width=\columnwidth]{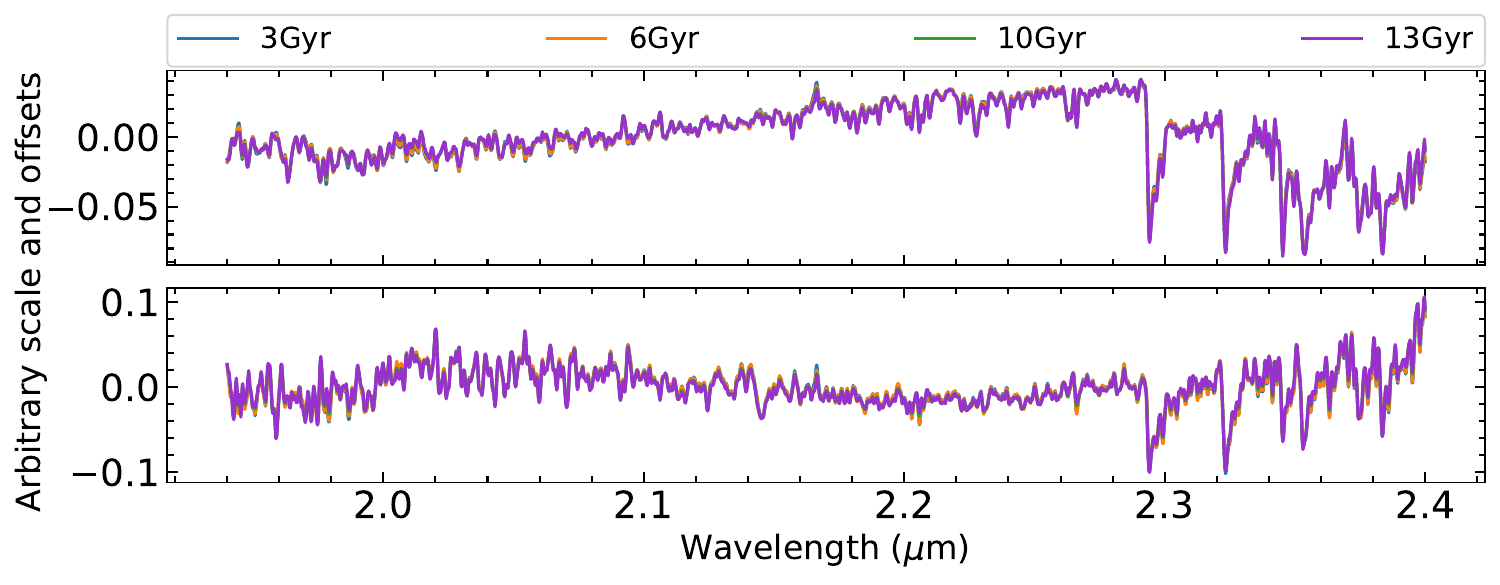}
 \caption{Same as Figure\,\ref{fig:age-variation_Iband} for the K-band.}
 \label{fig:age-variation_Kband}
\end{figure}

\begin{figure}
 \includegraphics[width=\columnwidth]{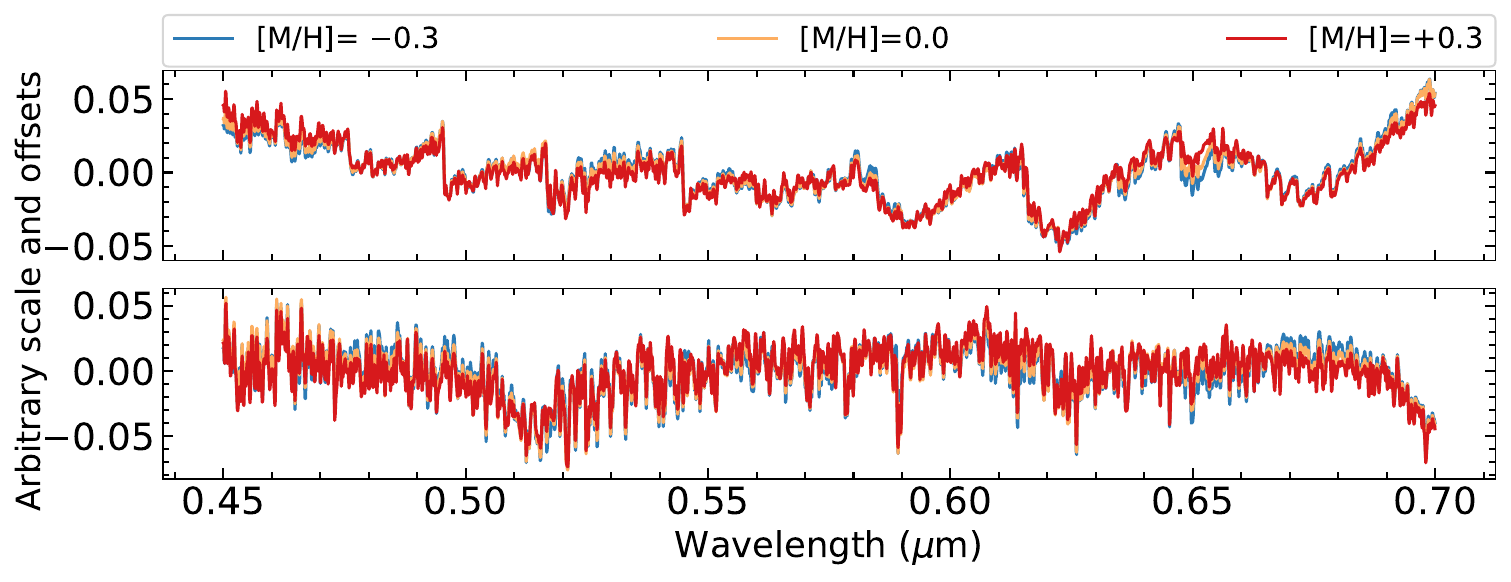}
 \caption{Same as Figure\,\ref{fig:feh-variation_Iband} for the V-band.}
 \label{fig:feh-variation_Vband}
\end{figure}

\begin{figure}
 \includegraphics[width=\columnwidth]{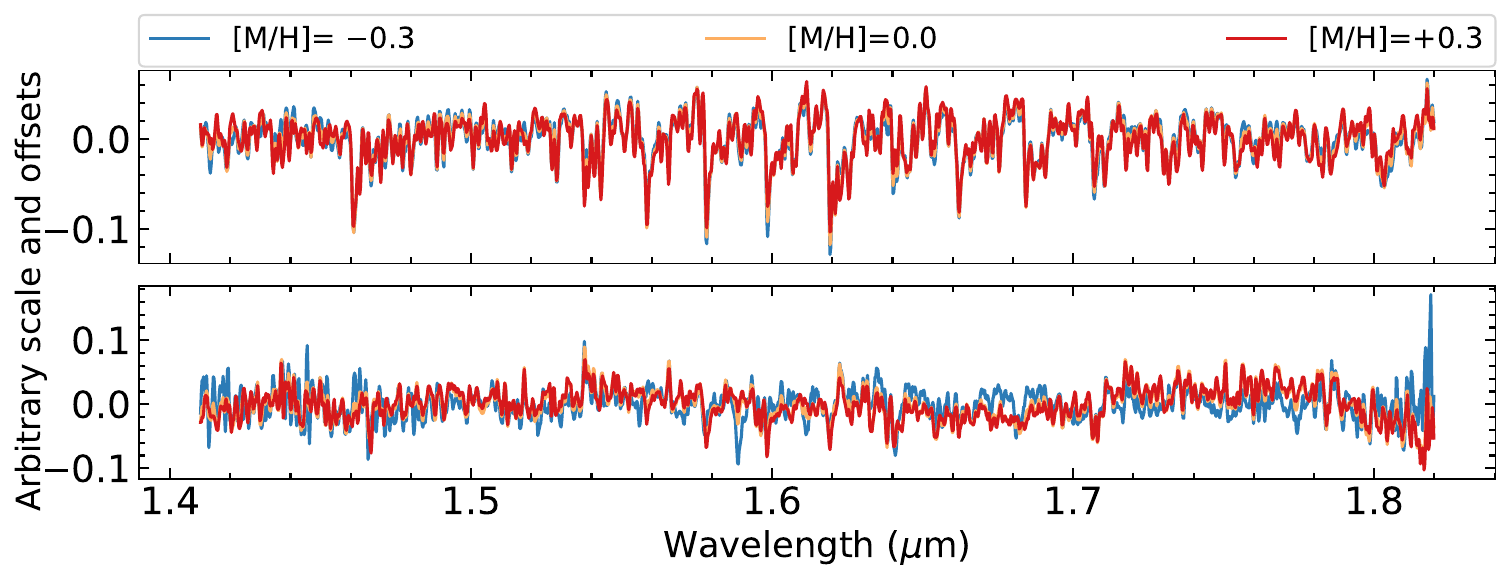}
 \caption{Same as Figure\,\ref{fig:feh-variation_Iband} for the H-band.}
 \label{fig:feh-variation_Hband}
\end{figure}

\begin{figure}
 \includegraphics[width=\columnwidth]{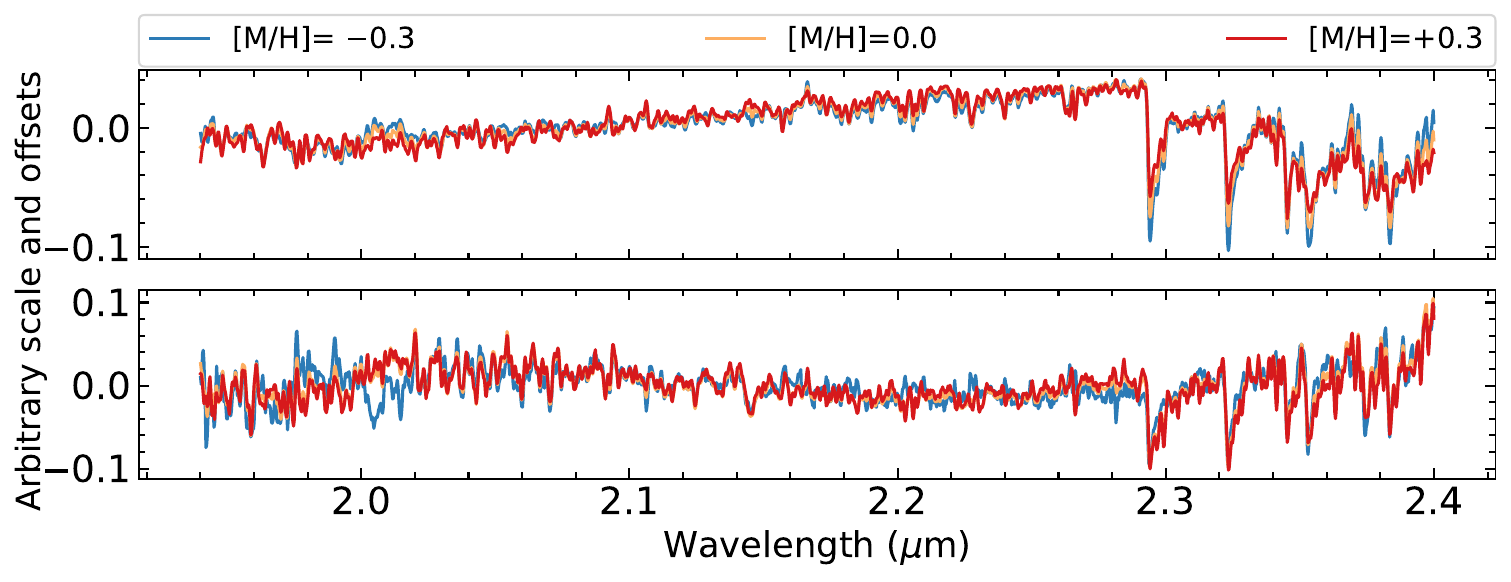}
 \caption{Same as Figure\,\ref{fig:feh-variation_Iband} for the K-band.}
 \label{fig:feh-variation_Kband}
\end{figure}

This appendix presents the complementary figures to those shown in the main text, completing the full set of results across all wavelength bands for the SSP cases spanning different ages, using solar-metallicity isochrones throughout (Figures\,\ref{fig:age-variation_Vband}, \ref{fig:age-variation_Jband}, \ref{fig:age-variation_Hband}, and \ref{fig:age-variation_Kband}), as well as the 10~Gyr cases covering a range of metallicities (Figures\,\ref{fig:feh-variation_Vband}, \ref{fig:feh-variation_Hband}, and \ref{fig:feh-variation_Kband}).

\section{COLIBRI isochrone extension}
\label{apx:colibri}
\begin{figure*} 
 \includegraphics[width=15cm]{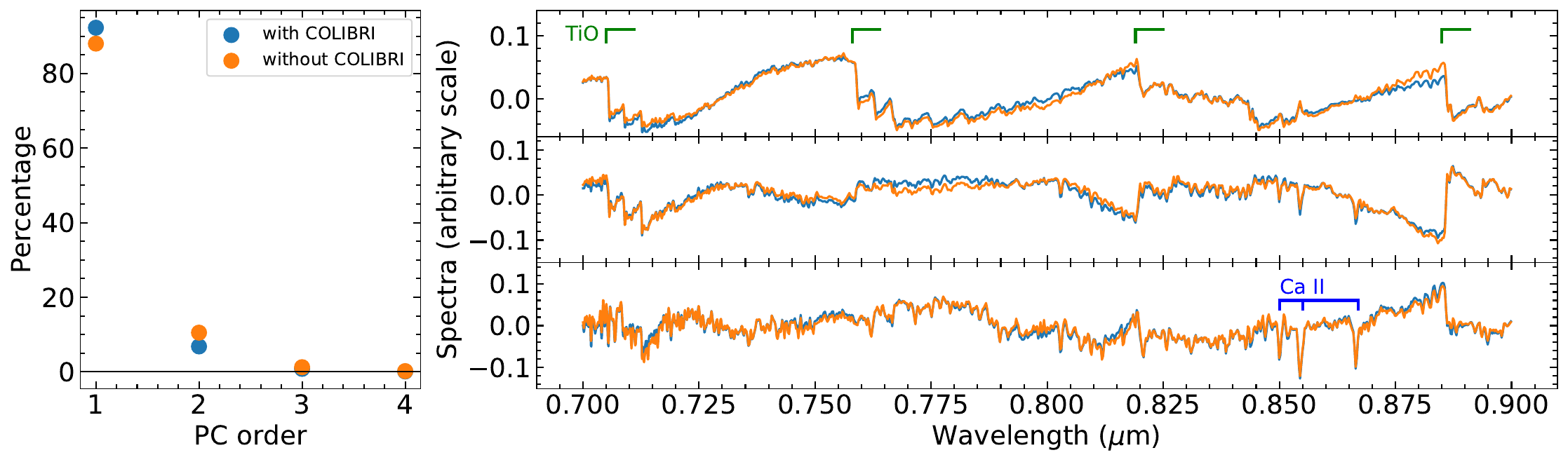}
 \caption{I-band eigenspectra and their contributions in percentages with and without COLIBRI addition, 10 Gyr, solar metallicity. Top: the proportion of total information contained in each of the first four PCs; blue shows the case with COLIBRI (0.93, 0.06, 0.007, 0.0006), orange - without the COLIBRI addition (0.88, 0.11, 0.01, 0.002). The second, third, and fourth plots display the first, second, and third eigenspectra, respectively, where blue shows the case with COLIBRY addition and the orange is without.}
 \label{fig:COLIBRI}
\end{figure*}

This appendix presents a comparison of the first three derived eigenspectra for isochrones with and without the inclusion of COLIBRI, as shown in Figure~\ref{fig:COLIBRI}. The inclusion of the COLIBRI extension modifies the stellar evolutionary tracks beyond the horizontal branch phase.
Up to the horizontal branch, the evolutionary tracks are effectively identical between the standard PARSEC models and those including COLIBRI. However, at later evolutionary stages, the stellar evolution is recomputed using the COLIBRI prescriptions, resulting in a modified asymptotic giant branch (AGB) evolution.
The differences between the standard PARSEC and PARSEC+COLIBRI models are not solely due to the inclusion of additional TP-AGB stars, but also arise from the altered morphology of the AGB evolution preceding the TP-AGB phase \citep{Marigo2017}.
The effect of including COLIBRI should be understood as a replacement and recalculation of the late stellar evolutionary phases, rather than a simple additive correction to the original PARSEC isochrones.



\bsp	
\label{lastpage}
\end{document}